\def\ind {\hangindent = 4mm\hangafter = 1\parindent = 0pt\parskip = 0pt}

\def\Msun{M$\sb{\odot}$}

\def\noy#1#2{\sp{#1}{\mathrm{#2}}}

\def\pa#1#2#3#4{\noy{#1}{#2}(\mathrm{p},\alpha)\noy{#3}{#4}}
\def\np#1#2#3#4{\noy{#1}{#2}(\mathrm{n,p})\noy{#3}{#4}}

\def\ag#1#2#3#4{\noy{#1}{#2}(\alpha,\gamma)\noy{#3}{#4}}

\def\FF{$\noy{19}{F}$}

\def\mass#1{${\mathrm{#1\,M}_\odot}$}

\def\chem#1#2{$\mathrm{^{#2}\kern-0.8pt#1}$}
\def\mchem#1#2{\mathrm{^{#2}\kern-0.8pt#1}}
\def\reac#1#2#3#4#5#6{$\mathrm{\,^{#2}\kern-0.8pt{#1}\,({#3}\,,{#4})\,{}^{#6}\kern-0.8pt{#5}\,}$}
\def\betap#1#2#3#4{$\mathrm{\,^{#2}\kern-0.8pt{#1}\,(\beta^+)\,{}^{#4}\kern-0.8pt{#3}\,}$}
\def\betam#1#2#3#4{$\mathrm{\,^{#2}\kern-0.8pt{#1}\,(\beta^-)\,{}^{#4}\kern-0.8pt{#3}\,}$}
\def\reacbp#1#2#3#4#5#6#7#8{$\mathrm{\,^{#2}\kern-0.8pt{#1}\,({#3}\,,{#4})\,{}^{#6}\kern-0.8pt{#5}\,(\beta^+)\,{}^{#8}\kern-0.8pt{#7}\,}$}
\def\reacbm#1#2#3#4#5#6#7#8{$\mathrm{\,^{#2}\kern-0.8pt{#1}\,({#3}\,,{#4})\,{}^{#6}\kern-0.8pt{#5}\,(\beta^-)\,{}^{#8}\kern-0.8pt{#7}\,}$}
\def\simgr{\mathbin{\;\raise1pt\hbox{$>$}\kern-8pt\lower3pt\hbox{$\sim$}\;}}
\def\simlr{\mathbin{\;\raise1pt\hbox{$<$}\kern-8pt\lower3pt\hbox{$\sim$}\;}}

\documentstyle[11pt,epsf]{article}

\begin{document}

\title{Fluorine production in intermediate-mass stars}
\markboth{Fluorine production in intermediate-mass stars}
{Fluorine production in intermediate-mass stars}

\author{ N. Mowlavi\ \thanks{Present Address: Observatoire de Gen\`eve, CH-1290 Sauverny, Switzerland} \and
         A. Jorissen\ \thanks{Research Associate, National Fund for Scientific Research (FNRS), Belgium} \and
         M. Arnould
       }

\date{
 \footnotesize Institut d'Astronomie et d'Astrophysique,\\
   \footnotesize  Universit\'e Libre de Bruxelles, C.P.226,\\
   \footnotesize  Boulevard du Triomphe,\\
   \footnotesize  B-1050 Bruxelles,\\
   \footnotesize  Belgium
}

\maketitle

\begin {abstract}
The \chem{F}{19} production during the first dozen thermal pulses of 
AGB stars with masses $M$ and metallicities $Z$ 
($M =$ \mass{3}, $Z =$ 0.02), ($M =$ \mass{6}, $Z =$ 0.02) and 
($M =$ \mass{3}, $Z =$ 0.001) is investigated on grounds of detailed
stellar models and of revised rates for \reac{N}{15}{\alpha}{\gamma}{F}{19}
and $\ag{18}{O}{22}{Ne}$. These calculations confirm an early expectation that
\chem{F}{19} {\it is} produced in AGB thermal pulses. They also enlarge
substantially these previous results by showing that the variations of the
level of \chem{F}{19} production during the evolution is very sensitive to the
maximum temperature reached at the base of the pulse. These variations are
analyzed in detail, and are shown to result from a subtle balance between
different nuclear effects (mainly \chem{F}{19} production or destruction in a
pulse, and \chem{N}{15} synthesis during the interpulse), possibly superimposed
on dilution effects in more or less extended pulse convective tongues. 

Our calculations, as most others, do not predict the third dredge-up 
self-consistently. When parametrized, it appears that our models 
of intermediate-mass AGB stars are able to
account only for the lowest \chem{F}{19} overabundances observed in 
solar-metallicity MS, S and C stars. That conclusion is expected to hold true for
low-mass stars when fluorine production results from secondary \chem{C}{13}
only. Massive AGB stars, on the other hand, are not expected to build up large
surface F abundances.  Therefore, the large fluorine overabundance reported for  
the super Li-rich star WZ~Cas (where hot bottom burning is supposed to be operating)
remains unexplained so far.
Our results for the (\mass{3}, $Z=0.001$) star indicate that fluorine surface
overabundances can also be expected in low-metallicity stars provided that
third dredge-ups occur after the early cool pulses. The relative increase in
the surface \chem{F}{19}/\chem{C}{12} ratio is, however, lower in the low-metallicity
than in the solar-metallicity star.  No observations are reported
yet for these stars, and are urgently called for.

\end{abstract}

\pagebreak

\section{Introduction}

Fluorine overabundances in giant stars
of spectral types MS, S and C have been reported recently (Jorissen, Smith 
\& Lambert 1992; Paper~I).
These observations reveal a progressive F enrichment of the envelope as the 
C/O ratio increases, reaching $[\mathrm{F/O}] \approx  1.5$ 
in some C stars. An independent indication of large F overabundances in carbon
stars comes from the recent detection of AlF in the inner envelope of the
dust-enshrouded carbon star IRC+10216 (Ziurys, Apponi \& Phillips 1994).
These observations shed light on the much debated question of the 
nucleosynthetic origin of \chem{F}{19}. They clearly point towards thermal
pulses as a likely site for its production.

The first quantitative proof that \chem{F}{19} can indeed be produced
in thermally pulsing AGB stars has been provided by Forestini et al. (1992;
Paper~II). However, this early investigation just considered the first 4
thermal pulses of a single (3~\Msun, $Z=0.03$) star. As a consequence, it left
unanswered several important questions, like the dependence of the level of
\chem{F}{19} production on the stellar mass, metallicity, or pulse number.

The present paper aims at exploring some of these questions. It analyzes a 
dozen pulses 
in three stars of different masses and metallicities ($M =$ \mass{3}, $ Z =$ 0.02;
$M =$ \mass{6}, $Z =$ 0.02; $M =$ \mass{3}, $Z =$ 0.001). Use is made of an improved 
code which follows more accurately the structural evolution of the AGB stars
and the concomitant nucleosynthesis. It also updates some nuclear reaction data
of importance in the study of the \chem{F}{19} production. In particular, a
more accurate rate (de Olivera et al. 1995) for the key reaction 
$\ag{15}{N}{19}{F}$ is adopted, as well as a recent re-evaluation (Giesen et 
al. 1994) of the $\ag{18}{O}{22}{Ne}$ rate. 
In addition, the surface fluorine enhancement due to the first and second
dredge-ups is also investigated.

The nuclear transformations leading to the production of $\noy{19}{F}$ in 
thermal pulses are reviewed in Sect.~\ref{Sect:nucleo}.
The evolution code and its key ingredients are briefly described in 
Sect.~\ref{Sect:ingredients},
while Sect.~\ref{Sect:predictions} presents our calculated \chem{F}{19} 
abundances. Our 
expectations are confronted with the observational data in 
Sect.~\ref{Sect:observations}, and 
conclusions are drawn in  Sect.~\ref{Sect:Conclusions}

\section{The nuclear transformations governing the \chem{F}{19} production 
in thermal pulses}
\label{Sect:nucleo}

\begin{figure}[t]
 \vskip 1.8cm
 \begin{picture} (4.7,4.7)
   \epsfysize=17.2cm
   \epsfxsize=17.2cm
   \epsfbox{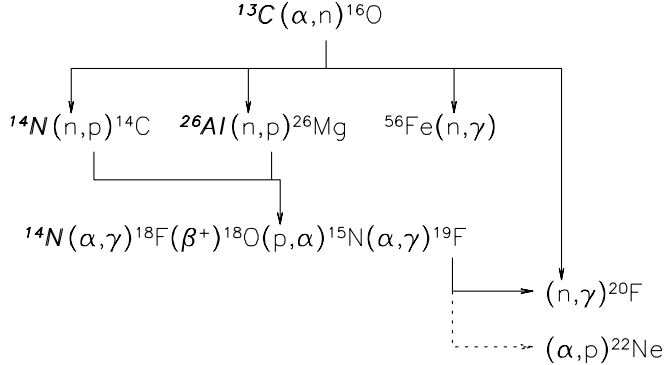}
 \end{picture}
 \vskip -1.5cm
\caption[]{
  \label{Fig:chain}
  The chain of reactions producing \chem{F}{19} in thermal pulse conditions.
 Bold italicized species are ashes from the H-burning shell. The most important
  (n,$\gamma$) capture competing with the indicated (n,p) reactions is
  \reac{Fe}{56}{n}{\gamma}{}{}
}
\end{figure}

The reaction chain producing \chem{F}{19} in thermal pulses has been identified in
Paper~II, and is displayed in Fig.~\ref{Fig:chain}. Its main 
characteristics may be summarized as follows:

\smallskip
\noindent (1) the main seeds for \chem{F}{19} production are \chem{C}{13} and 
\chem{N}{14} left behind by the H-burning shell. The resulting relatively small
amount of \chem{C}{13} is the main factor that limits the amount of \chem{F}{19}
possibly produced in a thermal pulse.  The efficiency of the \FF\ synthesis in
a pulse can be expressed as 
$(Y^*_{\mathrm{out}}(\mchem{N}{15})+Y^*_{\mathrm{out}}(\mchem{F}{19}))
 /Y_{\mathrm{in}}(\mchem{C}{13})$. In this
ratio, $Y_{\mathrm{in}}$(\chem{C}{13}) represents the abundance by number of
\chem{C}{13} injected in the pulse, and $Y^*_{\mathrm{out}}(\mchem{N}{15})$ and
$Y^*_{\mathrm{out}}(\mchem{F}{19})$ are the
abundances by number of \chem{N}{15} and \chem{F}{19} at the end
of the pulse,  assuming zero 
initial \chem{F}{19} and \chem{N}{15} abundances.

\smallskip
\noindent (2) the main loss of
efficiency in the \FF\ synthesis comes from the (n,$\gamma$) reactions, 
mainly on $\noy{56}{Fe}$ and heavier nuclei. They limit the occurrence of 
\reac{N}{14}{n}{p}{C}{14} and
\reac{Al}{26}{n}{p}{Mg}{26} reactions, which are the main producers of the
protons required for manufacturing \chem{N}{15}, the \chem{F}{19}
progenitor\footnote{The role of \chem{Al}{26} in enhancing the \chem{F}{19}
production in thermal pulses has been identified in Paper~II. The H-burning
shell of AGB stars is known to produce large amounts of \chem{Al}{26}
(Forestini et al. 1991)}.

\smallskip
\noindent (3) almost all the protons liberated as indicated above can lead to
the production of \FF\ if indeed the   
\reac{O}{18}{p}{\alpha}{N}{15}$(\alpha,\gamma)$\chem{F}{19} chain develops.
A necessary condition for this to occur is the availability of \chem{O}{18}.
Some \chem{O}{18} might be inherited from previous pulses.
As no interpulse \chem{O}{18} burning is predicted by our models,
this requires mainly that the duration $\Delta t_{\mathrm{pulse}}$ of 
these pulses be shorter than the destruction time scale of \chem{O}{18} against
$\alpha$-capture.
If this is not the case, no 
\chem{O}{18} from former pulses can survive at the beginning of a pulse.
In such conditions, and as 
\reacbp{N}{14}{\alpha}{\gamma}{F}{18}{O}{18} is slower than \reac{C}{13}{\alpha}{n}{O}{16}
for typical 
pulse temperatures, one would expect the neutrons to be liberated and captured
by the heavy nuclides before any \chem{O}{18} is made available. In this case,
the reaction chain leading the synthesis of 
\chem{F}{19} displayed in Fig. 1 would not operate.

In fact, the situation as it comes
out of the detailed calculations to be reported in Sect.~\ref{Sect:predictions}
is quite different. In reality, the nuclear sequence of Fig.~\ref{Fig:chain}
can still operate in a pulse {\it even if} that pulse does not inherit any
\chem{O}{18} from
previous pulses. We refer to Sect.~\ref{Sect:predictions} for details.

\smallskip
\noindent (4) \chem{F}{19} has the rather large Maxwellian-averaged
neutron-capture cross section of 5.7~mb at 30~keV (Beer, Voss \& Winters
1992). It has been checked that
\noindent (i) \reac{F}{19}{n}{\gamma}{F}{20} has no 
influence on the neutron budget. This results from the fact that \chem{N}{14}
or \chem{Fe}{56} are much more important neutron captors than \chem{F}{19}; and 
\noindent (ii) \reac{F}{19}{n}{\gamma}{F}{20} could
become a significant \FF\ destruction channel only for 
much larger neutron fluences than the ones predicted by the models reported 
in this paper.
In fact, if the larger neutron production would result from a primary \chem{C}{13}
reservoir, an {\it increased} \chem{F}{19} production would also be obtained from 
the chain of transformations of Fig. 1, thus limiting the impact of
\reac{F}{19}{n}{\gamma}{F}{20}. The models reported in this paper do not
consider such a possibility, which deserves further detailed studies.

\smallskip
\noindent (5) as shown in Fig. 1, \reac{F}{19}{\alpha}{p}{Ne}{22} could destroy some of
the produced \FF\ if the corresponding destruction time scale,
$\tau_{\alpha}$(\chem{F}{19}), gets 
shorter than $\Delta t_{\mathrm{pulse}}$. This happens when the
temperature at the base of the pulse exceeds typically $\sim 300\,10^6$ K.
The \FF\ production is therefore expected to be less efficient in the late 
AGB evolutionary stages, when the pulse temperatures are high.

\smallskip
Finally, it has to be noted that the \FF\ synthesis in convective 
pulses involves species that may have nuclear lifetimes shorter than, or 
similar to, the 
convective mixing time scale. This situation is encountered for neutrons and
protons. A proper treatment of the \FF\ production may 
therefore require the use of an algorithm coupling nucleosynthesis
and convective diffusion. This question will be addressed further in
Sect.~\ref{Sect:predictions}.

\section{Key ingredients of the stellar evolution code}
\label{Sect:ingredients}

A new implicit stellar evolution code that differs in several ways from the
one used in Paper II has been designed in order to follow
accurately the evolution of the structurally very complex AGB stars (see 
Mowlavi 1995a for a full description of the code). Its new features significantly
improve the numerical accuracy of the models, and avoid spurious numerical
effects exhibited by the models of Paper II.

In particular, the algorithm defining the discrete mass zones
is based on the structural equations, rather than on the relative
variations of the dependent variables. This provides a better handle on the
accuracy achieved on the structure\footnote{This structural accuracy is different
from the relative precision required for the convergence of
the Newton-Raphson solution of the structural equations. The latter
easily reaches $10^{-5}$ in all our models.}.
Convective zones growing into regions of variable
chemical composition have been modeled with great care. A special algorithm 
moreover ensures that composition discontinuities remain sharp as the star
evolves, thus avoiding ``numerical chemical diffusion'' as encountered in
the models of Paper II.

\begin{figure}
 \hskip -0.9cm
 \parbox[b]{13.3cm}{
 \vskip 1.3cm
 \begin{picture} (13.3,6.7)
   \epsfysize=19.5cm
   \epsfxsize=19.5cm
   \epsfbox{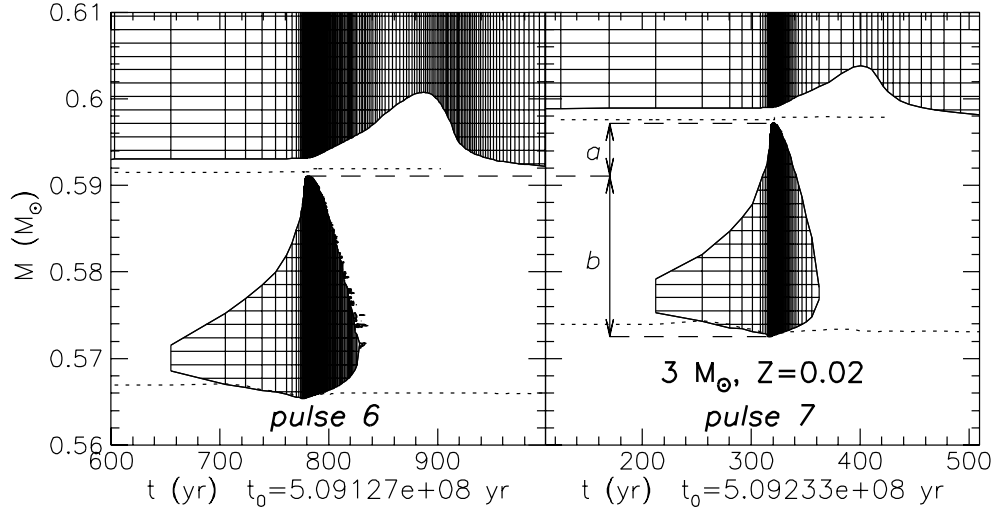}
 \end{picture}
 \vskip -1.8cm
 }
\hfill
\caption[]{ \label{Fig:pulseM3Z02}
  The 6th and 7th pulses of the 3~\Msun, $Z=0.02$ star. The scales
  are the same in both panels.  Hatched regions correspond to convective
  zones.  Each computed model is represented by a vertical line in
  the convective zones
  (horizontal lines have no special meaning).
  The short-dashed lines identify the location of maximum energy
  production in the H-burning (top) and He-burning (bottom) layers.
  Regions {\it a} and {\it b} as defined in the text are also indicated
  }
\end{figure}

 Each model is computed by iterating the
nucleosynthesis + mixing and the structure calculations. Starting
with an initial model at time $t$, taken identical to the model at time
$t-\Delta t$, where $\Delta t$ is the time step between the two models, a first
iteration is performed by calculating the nucleosynthesis + mixing, followed by
the calculation of the structure through a Newton-Raphson method. A second
iteration performs the nucleosynthesis + mixing on the last calculated
structure, followed by a new calculation of the structure. The procedure goes
on until the structure and the chemical distribution have converged. Two to
three such iterations are usually sufficient. From the core helium-burning phase
on, however, the number of iterations is limited to two, still providing a
satisfactory accuracy within reasonable computer times.
Rezoning is applied after each structure calculation in order to reach the required 
accuracy, and to predict the location of the convective 
boundaries to better than $10^{-5}$ \Msun.
Instantaneous mixing and concomitant homogeneous composition are assumed in 
the convective zones, except for nuclides with nuclear 
lifetimes $\tau_{\mathrm{nuc}}$ shorter than the convective mixing time
scale $\tau_{\mathrm{mix}}$. More specifically, the nucleosynthesis is calculated
in a one-zone approximation making use of an effective reaction rate obtained
by averaging the reaction rate over the convective zone
(see Prantzos, Arnould \& Arcoragi 1987 for more details).
When $\tau_{\mathrm{nuc}} < \tau_{\mathrm{mix}}$ for a given nuclide,
its abundance must be computed separately in each layer. In practice, it is
found that these nuclides (mainly neutrons, protons and sometimes \chem{O}{18})
usually also satisfy the condition of local equilibrium between production and
destruction, from which their abundance can easily be computed in each layer.
This procedure is only applied to nuclides with $\mathrm{X}\le 10^{-15}$ in the
convective zone (where X is the mass fraction), like neutrons and often
protons. Species more abundant than $\mathrm{X}=10^{-15}$ or with $\tau_{\mathrm{nuc}}
\sim\tau_{\mathrm{mix}}$ require an algorithm {\it coupling} nucleosynthesis and
convective diffusion. Such an algorithm has been implemented in our code, but
is very time-consuming. It has been applied to the calculation of one pulse, as
described in Sect.~\ref{Sect:M3Z02}, in order to evaluate the accuracy of the
results derived from the simpler procedure.

Nuclear reaction rates are taken from Caughlan \& Fowler 
(1988; CF88), except for the reactions listed in Appendix.
New reaction rates of importance for the \FF\ nucleosynthesis 
include $\ag{15}{N}{19}{F}$ (de Oliveira et al. 1995) and $\ag{18}{O}{22}{Ne}$
(Giesen et al. 1994). The new 
$\ag{15}{N}{19}{F}$ rate is slower than the CF88 one by a factor
1.5 to 50 in the temperature range $2 \le T_8 \le 3$ of interest 
here ($T_8$ is the temperature in units of $10^8$~K). The impact of this change
on the \FF\ nucleosynthesis has been investigated by de Oliveira 
et al. (1995).

\begin{figure}
 \hskip -0.9cm
 \parbox[b]{13.3cm}{
 \vskip 1.3cm
 \begin{picture} (13.3,7.5)
   \epsfysize=19.5cm
   \epsfxsize=19.5cm
   \epsfbox{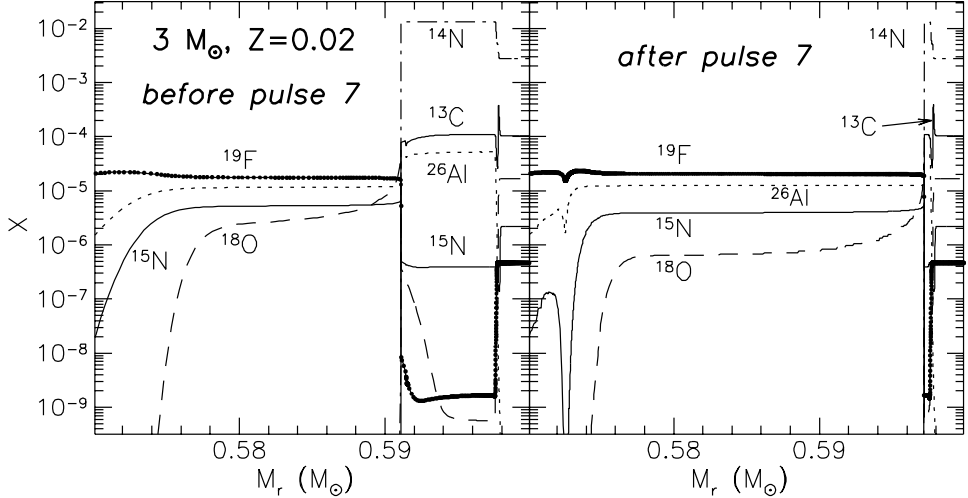}
 \end{picture}
 \vskip -1.8cm
 }
\hfill
\caption[]{ \label{Fig:profilesM3Z02}
Abundance profiles before (left panel), and after (right panel)
the 7th pulse of the \mass{3}, Z=0.02 star. Regions {\it a} and {\it b} defined
in the text are readily identified on the left panel. In the left panel, the
rapid drop of the
\chem{F}{19} abundance to the right of the plateau corresponds
to the maximum
outward extension of pulse 6, and marks the transition between regions
{\it a} and {\it b}. The dots on the \chem{F}{19}
profiles correspond to the mesh distribution
 }
\end{figure}

\section{Model predictions}
\label{Sect:predictions}

  Three intermediate-mass stars are computed with
(mass, metallicity) pairs of (3, 0.02), (3, 0.001) and (6, 0.02),
referred to as M3Z02, M3Z001 and M6Z02, respectively.

  Our models are computed from the pre-main sequence phase, when
the star is on the Hayashi track with central temperatures less
than $10^{6}$ K.
The structural and chemical evolution of the three stars are followed
all the way up to the 13th, 15th and 11th pulse of the
AGB phase for the M3Z02, M3Z001 and M6Z02 stars, respectively.
A detailed description of the
structural and chemical evolution of these stars can be found in Mowlavi
(1995a).  In the remaining part of this paper, we concentrate only on the
fluorine production in the AGB phase.

\subsection{The 3~\Msun, $Z = 0.02$ case}
\label{Sect:M3Z02}

The features governing the \chem{F}{19} synthesis in relatively cool thermal
pulses can be illustrated by considering the 6th and 7th pulses of the M3Z02
star. The location of the convective boundaries of these pulses is displayed in
Fig.~\ref{Fig:pulseM3Z02} as a function of time.
The pulse duration, $\Delta t_{\mathrm{pulse}}$,
is $182$ and 159 y for the 6th and 7th pulse, respectively,
and the maximum base temperature $T_{\mathrm{pulse,b}}=$ 239
and $245\,10^6$ K, respectively. Defining the maximum inward and outward
extensions of pulse $n$ as $M_{p,b}^{(n)}$ and $M_{p,t}^{(n)}$, respectively, the
overlap
\begin{displaymath}
  r=\frac{ M_{p,t}^{(6)}-M_{p,b}^{(7)} }{ M_{p,t}^{(7)}-M_{p,b}^{(7)} }
\end{displaymath}
between pulses 6 and 7 amounts to 0.753. Thus 75.3\% of the mass of
pulse 7 (defined as region {\it b} on
Fig.~\ref{Fig:pulseM3Z02}, and extending from $M_r=0.572$ to \mass{0.591})
is inherited
from pulse 6, the remaining fraction (defined as region {\it a} on
Fig.~\ref{Fig:pulseM3Z02}, and extending from $M_r=0.591$ to \mass{0.597}) containing
the ashes left behind by the hydrogen-burning shell. These two regions are
readily identified on the left panel of Fig.~\ref{Fig:profilesM3Z02}, which
displays the abundance profiles of \chem{C}{13}, \chem{N}{14}, \chem{N}{15},
\chem{O}{18},
\chem{F}{19} and \chem{Al}{26} before pulse 7. The drop in \chem{N}{15} and
\chem{O}{18} observed at the bottom of region {\it b} translates the fact that
$\alpha$-capture reactions continue to operate on these nuclides in the hottest
layers below pulse 6. A concomitant small hump in the fluorine abundance is also
observed in these layers. On the other hand, the non-homogeneous profile
exhibited by \chem{O}{18} in the upper part of region {\it b} corresponds to
the \chem{O}{18} abundance frozen, in a given layer,
at its value in the 6th pulse at the time the considered layer became radiative.

\begin{figure}[tbh]
 \hskip -0.8cm
 \parbox[b]{8.0cm}{
 \vskip 0.2cm
 \begin{picture} (8.0,9.6)
   \epsfysize=19.0cm
   \epsfxsize=19.0cm
   \epsfbox{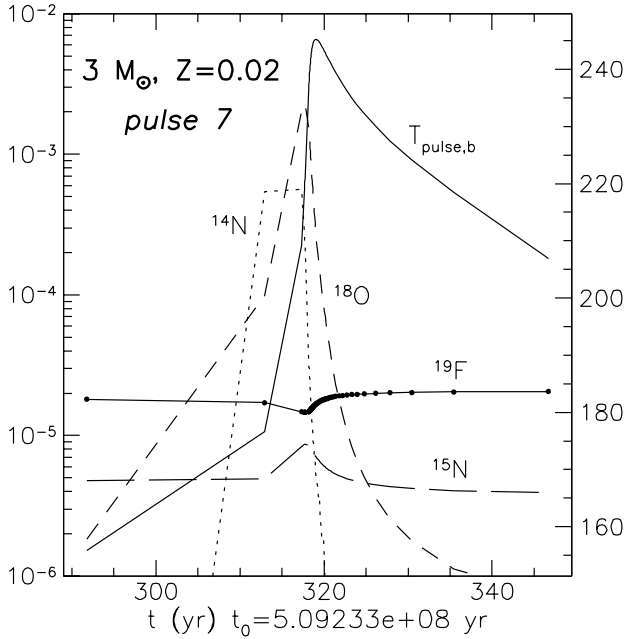}
 \end{picture}
 \vskip -1.3cm
 }
\caption[]{ \label{Fig:timeevolM3Z02}
  Evolution of the $\noy{14}{N}$, $\noy{15}{N}$, $\noy{18}{O}$ and 
  $\noy{19}{F}$ mass fractions during the 7th pulse of the \mass{3}, Z=0.02
  star. The evolution of the temperature at the
  base of the pulse is also shown (solid line), to be read on the right scale
  (in units of $10^6$~K). The quantities are plotted every 4 to 6 models,
  resulting in a poor time-resolution for some species
}
\end{figure}

  The evolution of the abundances of several nuclides of interest during the
7th pulse is shown in Fig.~\ref{Fig:timeevolM3Z02}.
The ingestion of \chem{N}{14} and the concomitant production of
\chem{O}{18} are clearly visible during the growing phase of the pulse.
The \chem{F}{19} abundance, on the other hand, first decreases slightly
as the pulse grows into fluorine-depleted regions.
It then increases as soon as \reac{N}{15}{\alpha}{\gamma}{F}{19} starts
operating.
A net increase in the fluorine abundance at
the end of the pulse, as compared to its value before the pulse, is clearly
apparent on Figs.~\ref{Fig:profilesM3Z02} and \ref{Fig:timeevolM3Z02},
even though the conversion of \chem{N}{15} into \chem{F}{19} has not been
complete. It is noteworthy that such a net \chem{F}{19} production has been
possible {\it even though} the amount of \chem{O}{18} inherited from the previous
pulses is about ten times lower than the \chem{F}{19} abundance level. In fact,
\chem{C}{13} and \chem{N}{14} are not ingested
instantaneously at the start of the pulse, but are instead mixed in
progressively. The ingestion time scale is set by the rate
of outward progression of the pulse convective tongue. This
time scale turns out to be longer than the one for \chem{O}{18} production. In
other
words, neutrons are released by \reac{C}{13}{\alpha}{n}{O}{16} in presence
of \chem{O}{18} produced earlier in the pulse by
\reac{N}{14}{\alpha}{\gamma}{F}{18}$(\beta^+)$\chem{O}{18} (see
Fig.~\ref{Fig:timeevolM3Z02}).
 
\begin{figure}[tbh]
 \hskip -1.5cm
 \parbox[b]{8.0cm}{
 \vskip 0.0cm
 \begin{picture} (8.0,11.5)
   \epsfysize=19.5cm
   \epsfxsize=19.5cm
   \epsfbox{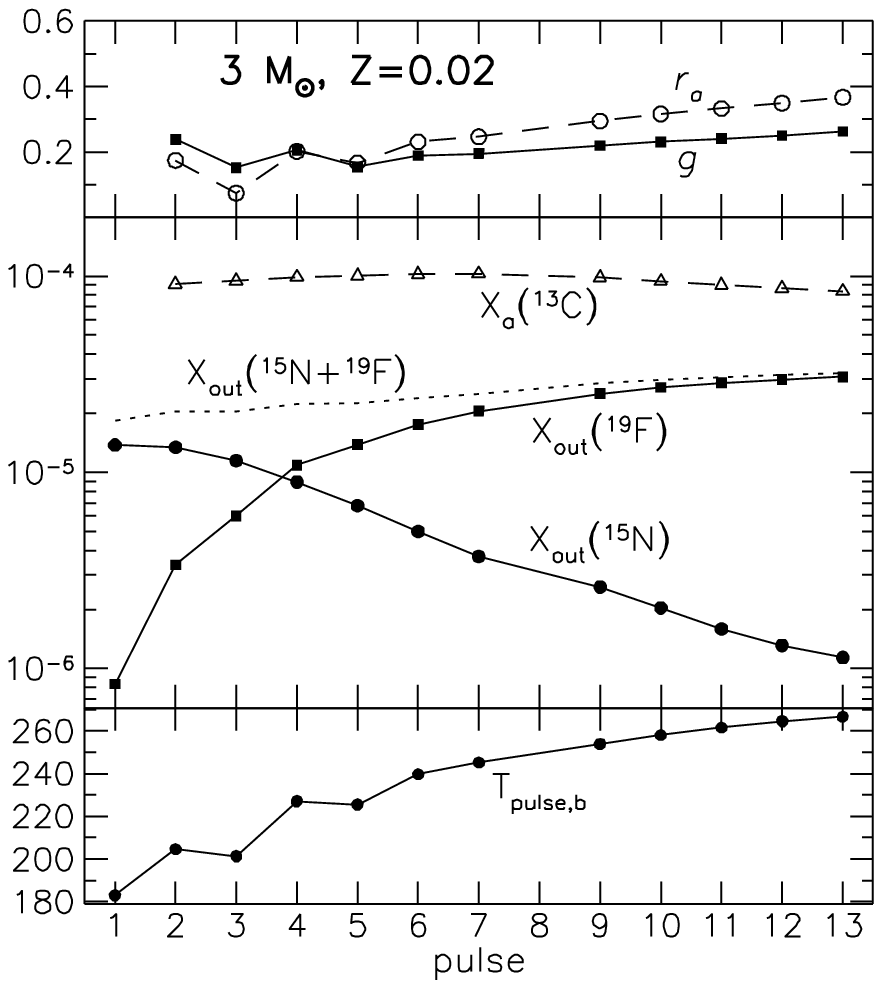}
 \end{picture}
 \vskip -1.8cm
 }
\caption[]{
\label{Fig:N15F19M3Z02}
Various quantities in the \mass{3}, Z=0.02 star as a function of pulse number.
{\it Lower panel}: Maximum temperature at the base of the thermal pulse;
{\it Middle panel}: Intershell \chem{N}{15} (filled circles), \chem{F}{19}
(filled squares) and \chem{N}{15}+\chem{F}{19} (dotted line) mass fractions at the
end of the pulse,
as well as \chem{C}{13} mass fraction averaged over region {\it a} as defined
in Fig.~\ref{Fig:pulseM3Z02} (open triangles); 
{\it Upper panel}: Efficiency factor $g$ of fluorine production in the pulse
(filled squares) and dilution factor $r_a$
of the hydrogen-burning ashes into the pulse (open circles)
}
\end{figure}

  The intershell mass fractions $X_{\mathrm{out}}$ of \chem{N}{15} and \chem{F}{19}
at the end of each
pulse are shown in Fig.~\ref{Fig:N15F19M3Z02} as a function of the pulse number.
The fraction of \chem{N}{15}
converted into \chem{F}{19} is low in the first pulses
\footnote{The conversion of \chem{N}{15} into \chem{F}{19} was faster with
the CF88 rate for \reac{N}{15}{\alpha}{\gamma}{F}{19}. We refer to de Oliveira
et al. (1995) for a discussion of the impact of the new rate on \chem{F}{19}
production in thermal pulses}.
It is, however, an increasing function of the pulse number, i.e. of the
maximum temperature reached at the base of the pulse.  The dotted line in
Fig.~\ref{Fig:N15F19M3Z02} represents the maximum \chem{F}{19} mass fraction
that would result from the complete conversion of \chem{N}{15} into
\chem{F}{19}. That limiting value is roughly constant with
pulse number. In other words, the efficiency of \chem{N}{15} production through
the above-mentioned reactions is not very sensitive to the pulse temperature,
at least when $T_{\mathrm{pulse,b}}<260\,10^6$ K.  The
fraction of \chem{N}{15} that can be converted into \chem{F}{19} in these
pulses {\it is} however sensitive to the
pulse temperature (Fig.~\ref{Fig:N15F19M3Z02}).

As already indicated in Sect.~\ref{Sect:nucleo}, the amount of \chem{N}{15}
that can be produced is limited by the available neutrons, i.e. by the 
available \chem{C}{13} supply. The maximum efficiency would 
correspond to a situation where each neutron liberated by
\reac{C}{13}{\alpha}{n}{O}{16}
produces a proton which is then captured by \reac{O}{18}{p}{\alpha}{N}{15}.
The actual efficiency can be
derived from baryon number conservation, as expressed by
the relation 
\begin{equation}
\label{Eq:eff}
g = \frac{\frac{19}{15} X_{\mathrm{out}}(\noy{15}{N}) + X_{\mathrm{out}}(\noy{19}{F})
        - \frac{19}{15} X_{\mathrm{in}}(\noy{15}{N})  - X_{\mathrm{in}} (\noy{19}{F})}
         {\frac{19}{13} X_{\mathrm{in}}(\noy{13}{C})},
\end{equation}
where the subscript {\it in} refers to the intershell
mass fractions averaged over region $a+b$ before the considered pulse,
and the subscript {\it out} to the abundance at the end of the pulse.
The average \chem{C}{13} mass fraction injected in the pulse, $X_{\mathrm{in}}
(\mchem{C}{13})$, results from the dilution into the pulse of \chem{C}{13}
left behind by the hydrogen-burning shell [$X_a(\mchem{C}{13})$ in the middle
panel of Fig.~\ref{Fig:N15F19M3Z02}].  The dilution factor
\begin{equation}
\label{Eq:dilution}
r_a=M_a/(M_a+M_b),
\end{equation}
where $M_a$ and $M_b$ are
the masses of regions {\it a} and {\it b}, respectively (see
Fig.~\ref{Fig:pulseM3Z02}), amounts to about 25\% for the 7th pulse. The actual
efficiency turns out to be around 20\%, as seen in the upper panel of
Fig.~\ref{Fig:N15F19M3Z02}.

  Another way of evaluating the efficiency is based on nuclear flux
considerations.  Leakage on the path from \chem{C}{13} to
\chem{N}{15} first occurs from neutrons liberated by
\reac{C}{13}{\alpha}{n}{O}{16} and involved in (n,$\gamma$) rather than in (n,p)
reactions. The main neutron poison in that respect is \chem{Fe}{56}.
The efficiency $g_1$ with which protons are liberated from a \chem{C}{13} seed is

\begin{eqnarray}
\label{Eq:g1}
 g_1 & = &
   \frac{ \sum\limits_{\mathrm{(n,p)}} \sigma_{\mathrm{A,np}} Y_{\mathrm{in}}(\mathrm{^AZ}) }
        { \left.\sum\limits_{\mathrm{n-capture}}\hspace{-3.8mm}\right.^*\hspace{1mm}
           \sigma_{\mathrm{A,n}} Y_{\mathrm{in}}(\mathrm{^AZ}) } \\
     & \simeq &
   \frac{ \sigma_{\mathrm{14,np}} Y_{\mathrm{in}}(\mchem{N}{14}) + 
          \sigma_{\mathrm{26,np}} Y_{\mathrm{in}}(\mchem{Al}{26}) }
        { \sigma_{\mathrm{14,np}} Y_{\mathrm{in}}(\mchem{N}{14}) + 
          \sigma_{\mathrm{26,np}} Y_{\mathrm{in}}(\mchem{Al}{26}) +   
          \sigma_{\mathrm{56,n}}  Y_{\mathrm{in}}(\mchem{Fe}{56}) }, \nonumber
\end{eqnarray}
where $Y(\mathrm{^AZ})=X(\mathrm{^AZ})/A$ and $\sigma$ stands for the Maxwellian-averaged
neutron-capture cross section.
The sum $\sum^*$ runs over all n-capture reactions except
\reac{C}{12}{n}{\gamma}{C}{13}, as this reaction replenishes \chem{C}{13}.
Since $\noy{14}{N}$ is only present in the hydrogen-burning ashes (region
{\it a}; see Fig.~\ref{Fig:profilesM3Z02}), 
its abundance after dilution in the pulse is
$X_{\mathrm{in}}(\mchem{N}{14}) = r_a X_a(\mchem{N}{14})$, with 
$X_a(\mchem{N}{14}) = 10^{-2}$, $X_a$ designating the
mass fraction averaged over region {\it a}.
A similar relation holds for \chem{Al}{26} in the M3Z02 star,
with $X_a(\mchem{Al}{26}) \sim 5\,10^{-5}$.
On the contrary, $X_{\mathrm{in}}(\mchem{Fe}{56})$ is independent of
$r_a$, as \chem{Fe}{56}
keeps its primordial abundance throughout the star. Therefore, 
as \chem{Fe}{56} constitutes the main neutron 
poison competing with $\np{14}{N}{14}{C}$ and $\np{26}{Al}{26}{Mg}$,
its importance will be comparatively larger, and thus $g_1$ smaller,
for smaller dilution factors $r_a$. {\it The efficiency of fluorine
production in the pulse is thus dependent on the pulse overlap factor $r=1-r_a$,
$g_1$ increasing with decreasing $r$}.
This is clearly seen on the upper panel of 
Fig.~\ref{Fig:N15F19M3Z02}, which shows that the efficiency correlates well
with the dilution factor.
On the contrary, the efficiency
is seen to be almost independent of the temperature, at least over the range 
$1.8 \le T_8 \le 2.5$. This is consistent with the fact that 
the Maxwellian-averaged rates for the neutron-capture reactions 
of interest are almost independent of temperature. 

  A second leakage in the fluorine production comes from protons captured by
nuclides other than \chem{O}{18}.
The efficiency $g_2$ with which protons liberated by (n,p) reactions contribute
to \reac{O}{18}{p}{\alpha}{N}{15} may be written
\begin{equation}
\label{Eq:g2}
  g_2=\frac{ \sigma_{\mathrm{18,p\alpha}} Y_{\mathrm{in}}(\mchem{O}{18}) }
           { \sum\limits_{\mathrm{p-capture}} \sigma_{\mathrm{A,p}} Y_{\mathrm{in}}(\mathrm{^AZ}) },
\end{equation}
where the symbols are as in Eq. (\ref{Eq:g1}).
The analysis of the reaction fluxes indicates that $g_2\sim 0.90$ in the pulses
of the M3Z02 star.

  Finally, the efficiency of fluorine production can be decreased further by
\reac{F}{19}{n}{\gamma}{F}{20} and/or \reac{F}{19}{\alpha}{p}{Ne}{22}. The
n-capture reaction does not play an important role given the low neutron
fluences encountered in our calculations.
The $\alpha$-capture reaction, on the other hand, is not
(yet) operating in the (still too cold ) pulses encountered in the M3Z02
star.  If $(1-g_3)$ is the fraction of \chem{F}{19} destroyed,
the total efficiency $g$ of fluorine production is given by
$g=g_1*g_2*g_3$.

  The maximum fluorine abundance that can be expected in the intershell region
can now easily be estimated from the above considerations. Assuming that
\chem{N}{15} is fully converted into \chem{F}{19} during the pulses (as is the
case after a few pulses), the \chem{F}{19} mass fraction at the end of pulse $n$
is equal to
\begin{equation}
\label{Eq:maxF19}
  X_{\mathrm{out}}^{(n)}(\mchem{F}{19})
   = (1-r_a) X_{\mathrm{out}}^{(n-1)}(\mchem{F}{19})
     + r_a \frac{19}{13} g X_a^{(n)}(\mchem{C}{13}),
\end{equation}
where the first term on the right-hand side represents the fluorine abundance
inherited from the
previous pulses, and the second term the amount of fluorine produced from the
\chem{C}{13} injected in the pulse. As fluorine builds up pulse
after pulse, its abundance will reach an asymptotic value when $X_{\mathrm{out}}^{(n)}
=X_{\mathrm{out}}^{(n-1)}$ (assuming a constant overlap factor).
Equation~(\ref{Eq:maxF19}) yields, in the asymptotic regime,
\begin{equation}
\label{Eq:F19asymptotic}
  X_{\mathrm{out}}(\mchem{F}{19}) = \frac{19}{13} g X_a(\mchem{C}{13})
\end{equation}
This value
can easily be computed from the equilibrium \chem{C}{13} mass fraction
$X_a(\mchem{C}{13})$
resulting from the CNO cycles, and from the neutron balance expressed 
by Eq.~(\ref{Eq:g1}). With $X_a(\mchem{C}{13})\simeq 10^{-4}$ and $g\simeq
0.20$, this leads to $X_{\mathrm{out}}(\mchem{F}{19})\simeq 3\,10^{-5}$, which is
precisely the asymptotic value obtained in the M3Z02 star
(Fig.~\ref{Fig:N15F19M3Z02}).

  The results presented above are obtained assuming instantaneous mixing in the
convective zone (except for neutrons). This assumption is, however, not
strictly valid for protons and in some cases \chem{O}{18}, whose nuclear
lifetimes may become shorter than the
convective mixing time scale. The use of a very computer time-consuming
algorithm coupling diffusion and nucleosynthesis (Sect.~\ref{Sect:ingredients}) does
not lead, however, to significantly different \chem{F}{19} abundances in a
test run on the 7th pulse.

\subsection{The 3~\Msun, $Z = 0.001$ case}
\label{Sect:M3Z001}

  The evolution of the fluorine abundance in the \mass{3}, Z=0.001
star is shown in Fig.~\ref{Fig:N15F19M3Z001} as a function of the
pulse number. Several important differences exist as compared to the
M3Z02 case:

\begin{figure}[tbh]
 \hskip -1.4cm
 \parbox[b]{8.0cm}{
 \vskip 0.0cm
 \begin{picture} (8.0,11.5)
   \epsfysize=19.5cm
   \epsfxsize=19.5cm
   \epsfbox{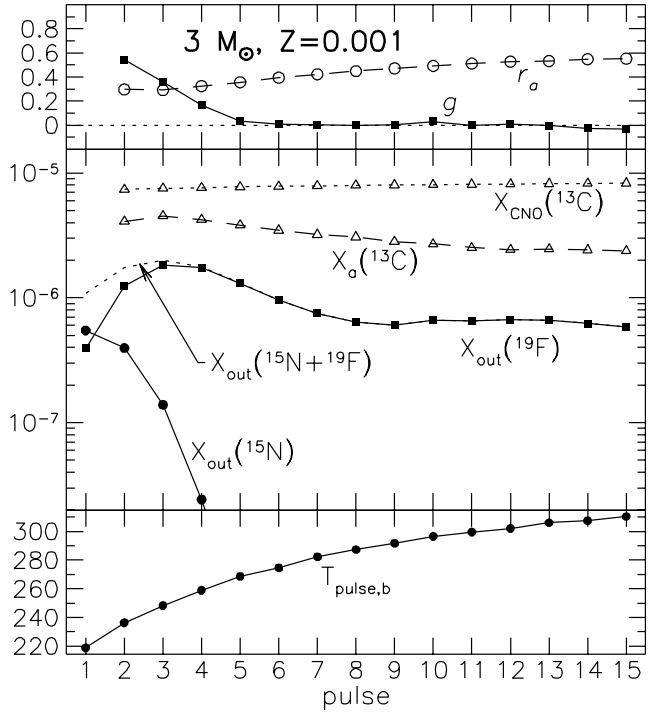}
 \end{picture}
 \vskip -1.8cm
 }
\caption{
\label{Fig:N15F19M3Z001}
Same as Fig.~\protect\ref{Fig:N15F19M3Z02}, but for the \mass{3}, Z=0.001 star. In the
middle panel, $\mathrm{X_{CNO}(\mchem{C}{13})}$ corresponds to the \chem{C}{13}
mass fraction left behind by the hydrogen-burning shell.
The difference between $\mathrm{X_{CNO}}(\mchem{C}{13})$ and $\mathrm{X_a}(\mchem{C}{13})$
is due to the interpulse \chem{C}{13} burning
}
\end{figure}

\smallskip
\noindent i)  The maximum fluorine abundance, corresponding to
the complete transformation of \chem{N}{15} into \chem{F}{19}, is
reached {\it much} faster in the M3Z001 star (after 3 pulses) than in
the M3Z02 star (at least 10 pulses are needed). This is due to the
higher temperatures reached in the pulses of the low-metallicity star.

\smallskip
\noindent ii) The efficiency of \chem{F}{19} synthesis (upper panel of
Fig.~\ref{Fig:N15F19M3Z001}) during the first pulses is much larger than in
the M3Z02 star ($g\sim 0.50$ as compared to 0.20). This is due to the combined
effect of smaller overlap factors between successive pulses (see the discussion
of Sect.~\ref{Sect:M3Z02}) and of larger \chem{Al}{26}/\chem{Fe}{56} abundance
ratios left behind by the hydrogen-burning shell and ingested by the pulse.
Arnould \& Mowlavi (1993, see also Arnould et al. 1995) showed that the \chem{Al}{26} abundance
resulting from hydrogen burning sensitively depends upon the temperature, being
maximum for temperatures around $T_6\sim 70$.
The temperature $T_{\mathrm{HBS}}$ in the hydrogen-burning shell is
indeed more favorable to \chem{Al}{26} production
in the M3Z001 star ($T_{\mathrm{HBS}}=73\,10^6$~K
before the first pulse) than in the M3Z02 star ($T_{\mathrm{HBS}}=45\,10^6$~K before the
first pulse). As a consequence, the ability of \chem{N}{14} and \chem{Al}{26}
to compete with the neutron poisons is much higher during the first pulses of
the M3Z001 star. In the
second pulse of the M3Z001 star, about 10 and 50\% of the neutrons liberated by
\reac{C}{13}{\alpha}{n}{O}{16} are captured by
\reac{N}{14}{n}{p}{C}{14} and \reac{Al}{26}{n}{p}{Mg}{26}, respectively.

\begin{figure}[tbh]
 \hskip -1.3cm
 \parbox[b]{8.0cm}{
 \vskip 0.6cm
 \begin{picture} (8.0,9.0)
   \epsfysize=22.0cm
   \epsfxsize=22.0cm
   \epsfbox{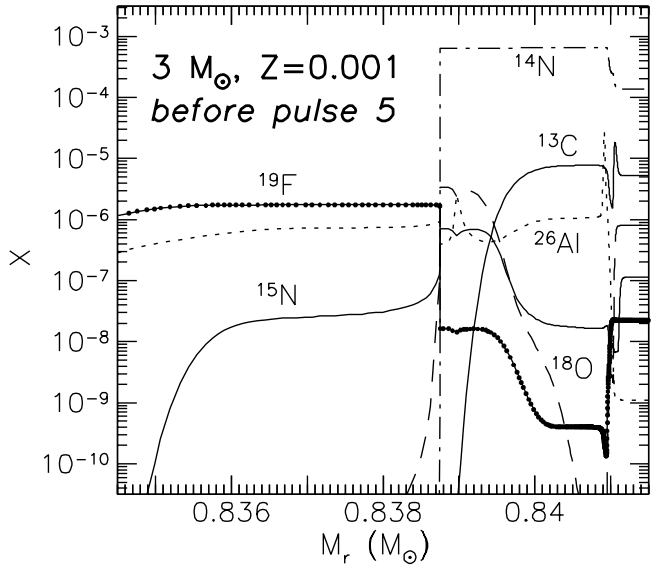}
 \end{picture}
 \vskip -2.0cm
 }
\caption[]{
\label{Fig:profilesM3Z001}
Same as Fig.~\protect\ref{Fig:profilesM3Z02}, but for the 5th pulse of the 
\mass{3}, Z=0.001 star
}
\end{figure}

\begin{figure}[tbh]
 \hskip -1.6cm
 \parbox[b]{8.0cm}{
 \vskip -0.1cm
 \begin{picture} (8.0,11.0)
   \epsfysize=19.5cm
   \epsfxsize=19.5cm
   \epsfbox{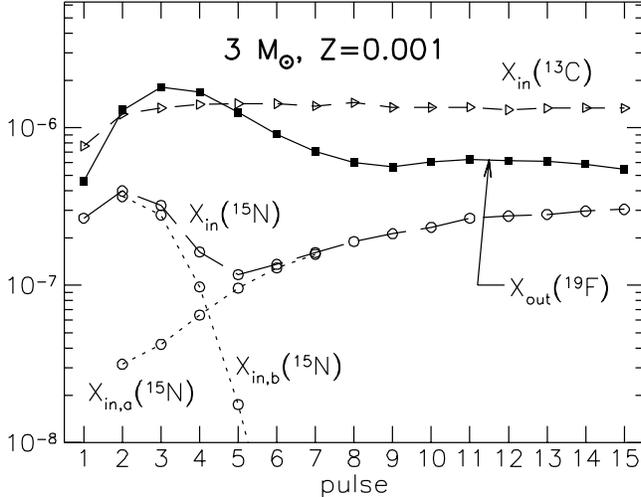}
 \end{picture}
 \vskip -4.5cm
 }
\caption[]{
 \label{Fig:N15init}
 Mass fractions $X_{\mathrm{in}}$, averaged over the total mass covered by the
pulse, of \chem{C}{13} (open triangles) and
 \chem{N}{15} (open circles, long-dashed line) before each pulse
 of the \mass{3}, Z=0.001 star.
 The contributions from regions $a$ and $b$ (see Fig.~\ref{Fig:pulseM3Z02}),
 to the \chem{N}{15} injected in the pulse are represented by open circles on
 doted lines, as labeled in the figure. The fluorine mass fraction $X_{\mathrm{out}}$
 after each pulse, averaged over the mass of the pulse, is given by the filled squares
}
\end{figure}

\begin{figure}[tbh]
 \hskip -0.7cm
 \parbox[b]{8.0cm}{
 \vskip 0.2cm
 \begin{picture} (8.0,9.6)
   \epsfysize=19.0cm
   \epsfxsize=19.0cm
   \epsfbox{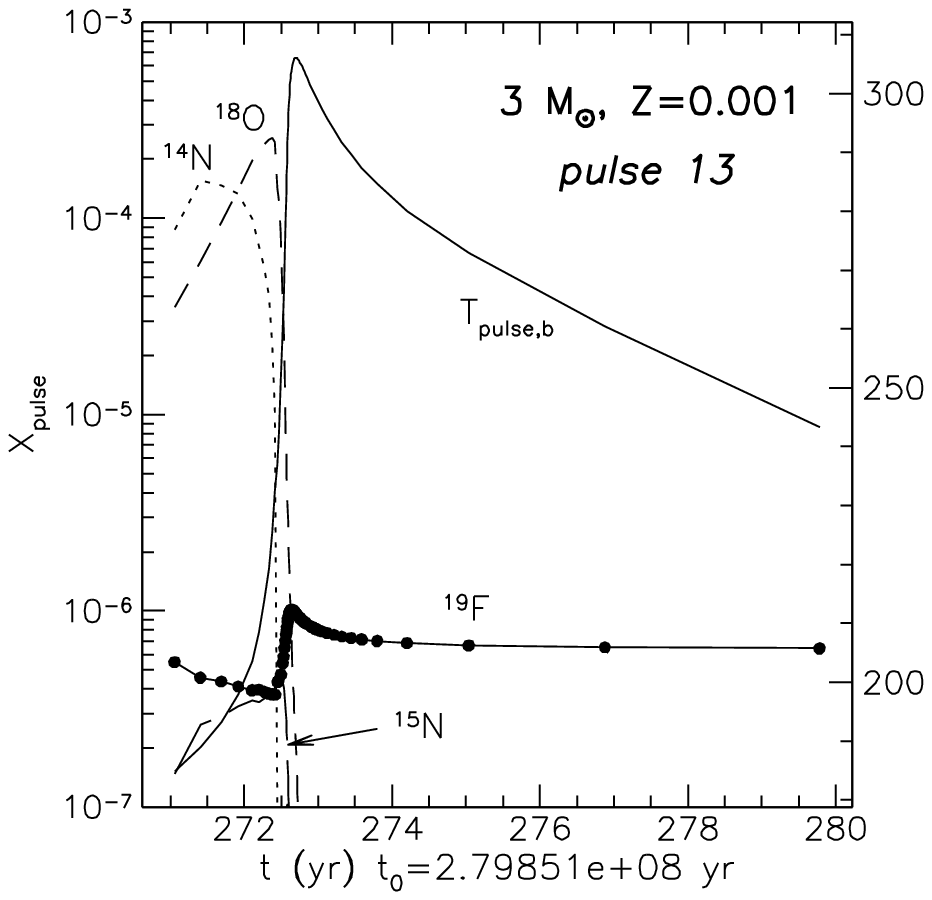}
 \end{picture}
 \vskip -1.4cm
 }
\caption{
\label{Fig:timeevolM3Z001}
Same as Fig.~\protect\ref{Fig:timeevolM3Z02}, but for the 13th pulse of
the \mass{3}, Z=0.001 star
}
\end{figure}

The efficiency, however, rapidly decreases, and drops to zero after the 5th
pulse. This results from two effects. First, the intershell \chem{Al}{26} abundance
decreases as the temperature of the hydrogen-burning shell exceeds
$70\,10^6$~K. Secondly, the temperature in the He-rich layers is high enough 
for \reac{C}{13}{\alpha}{n}{O}{16} to {\it destroy $^{\mathit{13}}\!$C in the deep layers
of region $a$ during the interpulse period}.
This is clearly seen on Fig.~\ref{Fig:profilesM3Z001}, which displays
the abundance profiles before the 5th pulse of the M3Z001 star. As a
consequence, the surviving \chem{C}{13} is injected in the pulse at a time where
the base temperature is close to its maximum.
At these high temperatures, the
destruction time scale of \chem{N}{14} against $\alpha$-capture
becomes smaller than its injection rate in the pulse, so that its abundance
decreases. Both effects drastically reduce the
ability of \chem{N}{14} and \chem{Al}{26} to compete with
\chem{Fe}{56} for capturing neutrons, resulting in a drop of the
efficiency of fluorine production within the thermal pulses.

\begin{figure}[tbh]
 \hskip -1.6cm
 \parbox[b]{8.0cm}{
 \vskip 0.0cm
 \begin{picture} (8.0,11.5)
   \epsfysize=19.5cm
   \epsfxsize=19.5cm
   \epsfbox{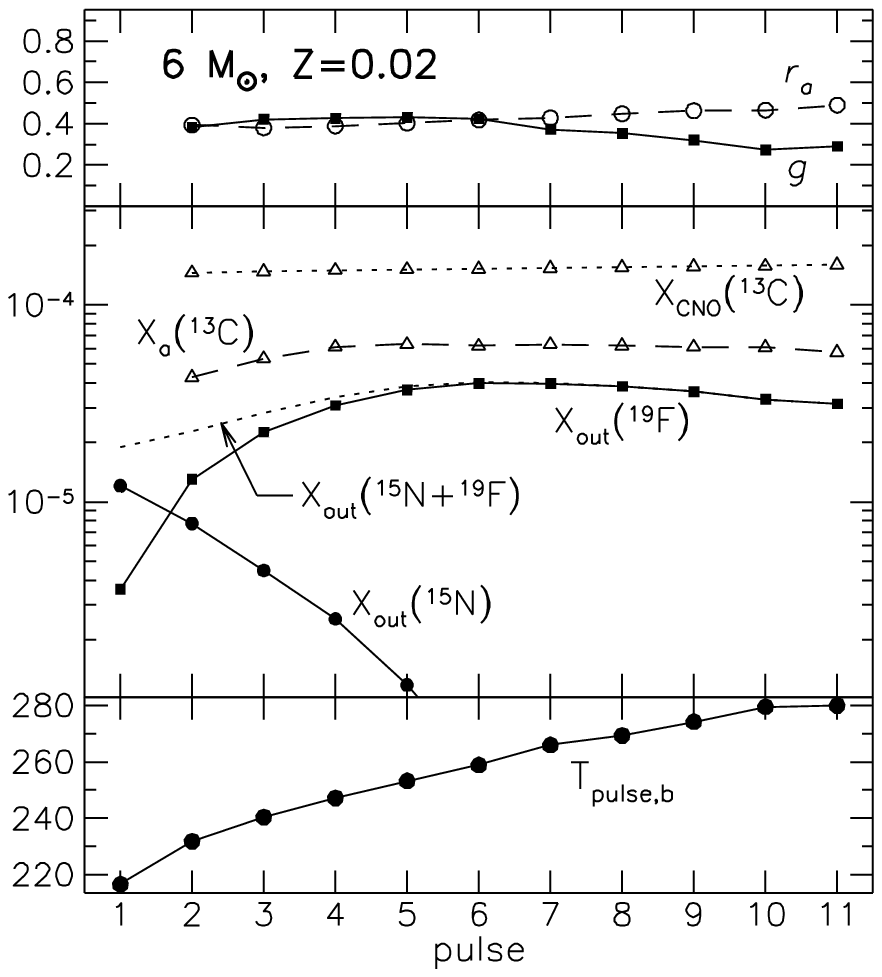}
 \end{picture}
 \vskip -2.0cm
 }
\caption{
\label{Fig:N15F19M6Z02}
Same as Fig.~\protect\ref{Fig:N15F19M3Z001}, but for the \mass{6}, Z=0.02 star
}
\end{figure}

\smallskip
\noindent iii) The interpulse \chem{C}{13} burning leads to
 concomitant {\it $^{\mathit{15}}\!$N production  in radiative layers},
as seen
on Fig.~\ref{Fig:profilesM3Z001}. When injected in the pulse, this \chem{N}{15}
will {\it directly} produce \chem{F}{19} through
\reac{N}{15}{\alpha}{\gamma}{F}{19}. Figure~\ref{Fig:N15init} shows how the
different contributions to the after-pulse fluorine abundance vary with pulse number.
In the early pulses, \chem{N}{15} is produced from \chem{C}{13} within the pulse,
with an efficiency decreasing with the pulse number.
The after-pulse fluorine abundance, resulting from the
conversion of \chem{N}{15}, reaches a maximum at pulse 3 and then decreases. From
pulse 5 on, a substantial amount of \chem{N}{15} is produced in region {\it a}
during the interpulse period [$X_{\mathrm{in,a}}(\mchem{N}{15})$], and represents,
after injection in the pulse [$X_{\mathrm{in}}(\mchem{N}{15})$],
the main contribution to the after-pulse fluorine
abundance. It is noteworthy that the F plateau reached with this radiative
\chem{N}{15} contribution is lower than the one that would have resulted
had \chem{N}{15} been synthesized in the pulse. This translates the fact
that the efficiency $g_{rad} (\sim 0.09)$ of \chem{N}{15} production in radiative
layers is actually lower than in convective (cold) pulses. Indeed, in a radiative
layer, the \chem{O}{18} necessary to produce \chem{N}{15} through
\reac{O}{18}{p}{\alpha}{N}{15} becomes available through
\reac{N}{14}{\alpha}{\gamma}{F}{18}($\beta^+$)\chem{O}{18},
only after a substantial amount of
\chem{C}{13} has already been burned. This results from the longer
\reac{N}{14}{\alpha}{\gamma}{F}{18}($\beta^+$)\chem{O}{18} time scale as
compared to the \reac{C}{13}{\alpha}{n}{O}{16} one.

\smallskip
\noindent iv) Finally, fluorine destruction through \reac{F}{19}{\alpha}{p}{Ne}{22}
begins to operate when the temperature at the base of the pulse exceeds
$300\,10^6$ K. This happens
from the 13th pulse on in the M3Z001 star, and is clearly visible in
Fig.~\ref{Fig:timeevolM3Z001}, which shows the time evolution of the abundances
in the 13th pulse. The efficiency of fluorine production
decreases to {\it negative} values as a result of the fluorine destruction
in the hot pulses of the low-metallicity star (Fig.~\ref{Fig:N15F19M3Z001},
upper panel).  The concomitant decrease of the after-pulse \chem{F}{19} abundance
is also observed in Fig.~\ref{Fig:N15F19M3Z001} (middle panel).

\subsection{The 6~\Msun, $Z = 0.02$ case}
\label{Sect:M6Z02}

\begin{figure}[tbh]
 \hskip -0.9cm
 \parbox[b]{8.0cm}{
 \vskip 0.8cm
 \begin{picture} (8.0,7.4)
   \epsfysize=18.5cm
   \epsfxsize=18.5cm
   \epsfbox{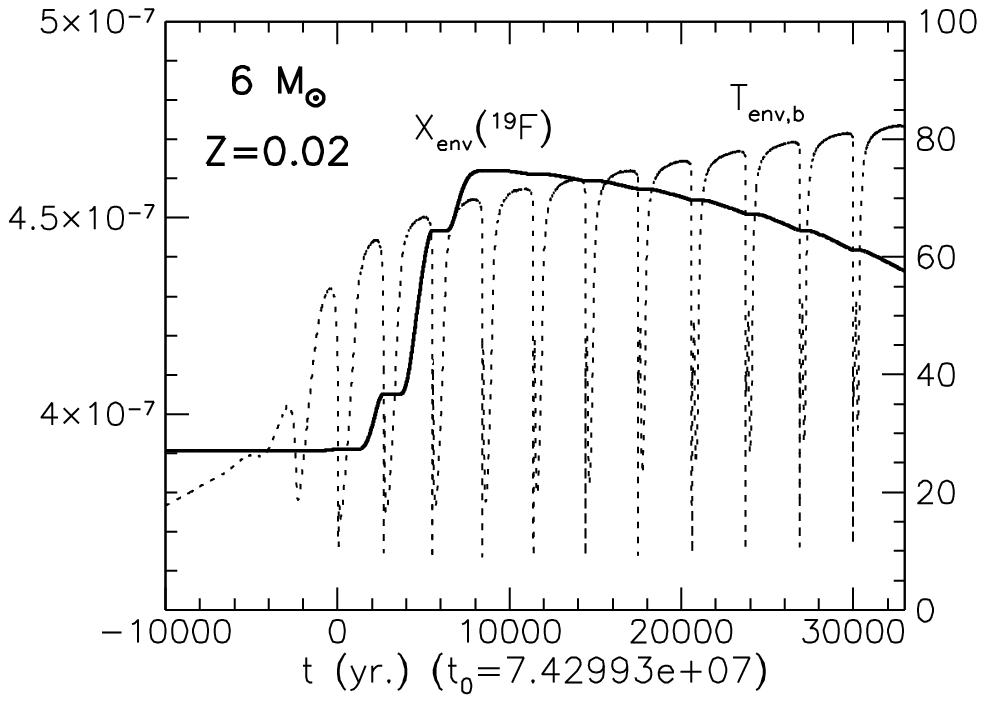}
 \end{picture}
 \vskip -1.6cm
 }
\caption{
\label{Fig:s6z02mHBBF19}
  Fluorine abundance (thick solid line, corresponding to the left scale) in the
envelope of the \mass{6}, Z=0.02 star, and temperature (dotted line, to be read
on the right scale, in units of $10^6$ K) at the base of the
envelope. Time $t=0$ corresponds to the occurrence of the first
convective pulse. The temperature drop after each pulse results from
the expansion, and concomitant cooling, of the hydrogen-rich layers following
the thermal pulses
}
\end{figure}

The situation for the M6Z02 star is qualitatively 
similar to that described for the M3Z001 star. The evolution of the after-pulse
\chem{F}{19} is simply delayed in the M6Z02 star as compared to the M3Z001
star, as a result of the lower pulse temperatures (namely $T_6=260$ and 290 in
the 8th pulse of the M6Z02 and M3Z001 stars, respectively).

  A specific feature of the M6Z02 star concerns however the temperatures at the base
of the convective envelope, $T_{\mathrm{env,b}}$, which are {\it much} higher in the
M6Z02 than in the M3Z02 and M3Z001 stars. At the first pulse, $T_{\mathrm{env,b}}$ amounts
to 2.5, 6 and $50\,10^6$ K in the M3Z02, M3Z001 and M6Z02 stars, respectively, and
to 4.6, 25 and $79\,10^6$~K at the 10th pulse, respectively. As a consequence,
hydrogen burning occurs {\it in} the envelope of the M6Z02 star. This feature
is known as hot-bottom burning (HBB).

\begin{figure}[tbh]
 \hskip -1.4cm
 \parbox[b]{8.0cm}{
 \vskip 0.7cm
 \begin{picture} (8.0,10.4)
   \epsfysize=19.5cm
   \epsfxsize=19.5cm
   \epsfbox{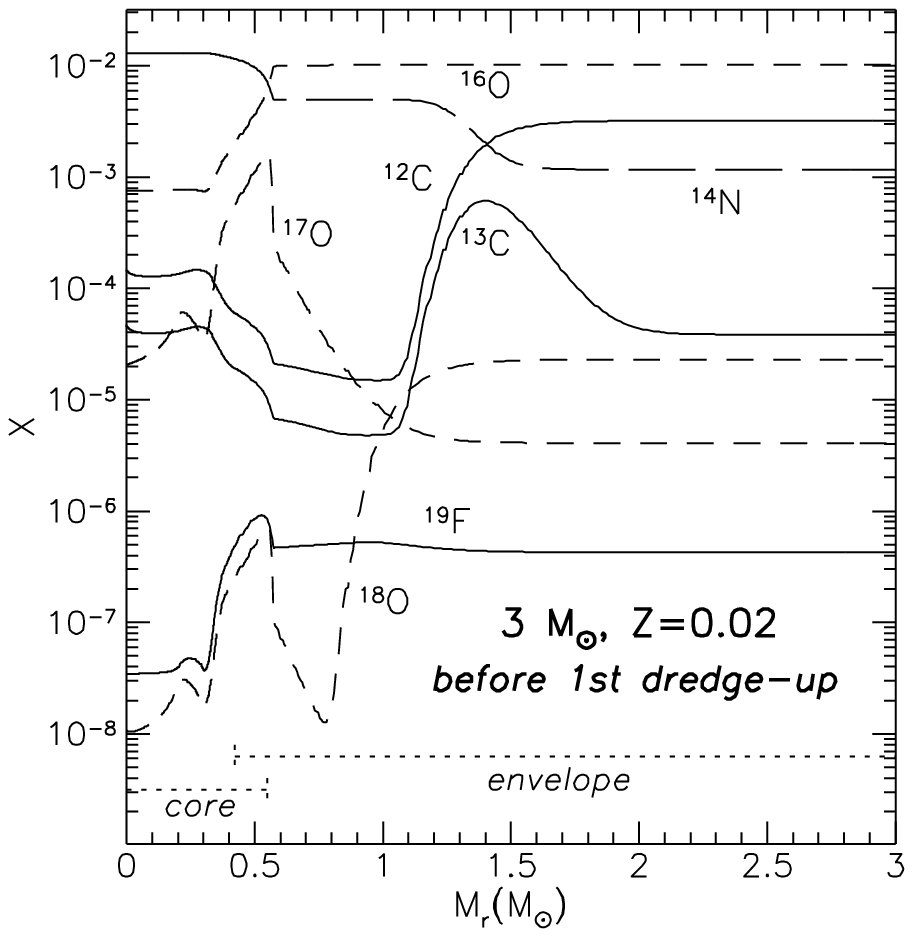}
 \end{picture}
 \vskip -1.6cm
 }
\caption[]{
\label{Fig:firstdredgeupprofiles}
Mass fraction profiles of \chem{C}{12}, \chem{C}{13}, \chem{N}{14}, \chem{O}{16},
\chem{O}{17}, \chem{O}{18} and \chem{F}{19} before the first dredge-up in the
M3Z02 star. The maximum extension of the convective core during central hydrogen burning
(labeled {\it core}), and the maximum inward penetration of the envelope during
the red giant phase (labeled {\it envelope}) are indicated at the bottom of the
figure
}
\end{figure}

  The effect of hydrogen burning on fluorine depends on the
temperature, as shown by Arnould \& Mowlavi
(1993). For $T_6<20$, fluorine is {\it produced} by \reac{O}{18}{p}{\gamma}{F}{19},
by a factor of up to 3 with respect to its solar-system abundance.
At higher temperatures, it is essentially {\it destroyed} by
\reac{F}{19}{p}{\alpha}{O}{16}. As a consequence of HBB, the \chem{F}{19} abundance in
the {\it envelope} of the M6Z02 star is thus expected
to first increase, and then to decrease as the temperatures get higher.
Figure~\ref{Fig:s6z02mHBBF19} shows that the maximum envelope \chem{F}{19} abundance
is reached when $T_{\mathrm{env,b}}=70\,10^6$~K~\footnote{The ``effective''
temperature characterizing
the nucleosynthesis in a convective zone is lower than the temperature at the
base of that zone, because temperature-dependent reaction rates must be averaged
over the whole convective region (see Sect.~\ref{Sect:ingredients})}, though the
overabundance remains small
([\chem{F}{19}]=log($X_{\mathrm{env}}(\mchem{F}{19})/X_\odot(\mchem{F}{19}))=0.07$).
More important
is the fluorine destruction in the envelope as the temperature
exceeds $T_6=70$, which happens from the 4th pulse on in the M6Z02 star.
Some fraction of the fluorine brought in the envelope by the third dredge-up
(which we did not obtain in a self-consistent way in our models; see
Sect.~\ref{Sect:thirddredge-up}) would thus be destroyed as a result of
the HBB.

\section{Comparison with observations}
\label{Sect:observations}

  Observations of fluorine abundances at the surface of giant stars are
available only for stars with metallicities close to solar (Paper I), as displayed on
Figs.~\ref{Fig:FOCOno3dup} and \ref{Fig:FOCO}.  We thus restrict the comparison
with observations to the M3Z02 and M6Z02 stars. Predictions for the low-metallicity
star are nevertheless briefly presented in Sect.~\ref{Sect:lowmetallicity}.

\subsection{First and second dredge-ups}
\label{Sect:firstdredge-up}

\begin{figure}[tbh]
 \hskip -1.3cm
 \parbox[b]{8.0cm}{
 \vskip 0.3cm
 \begin{picture} (8.0,9.0)
   \epsfysize=19.2cm
   \epsfxsize=19.2cm
   \epsfbox{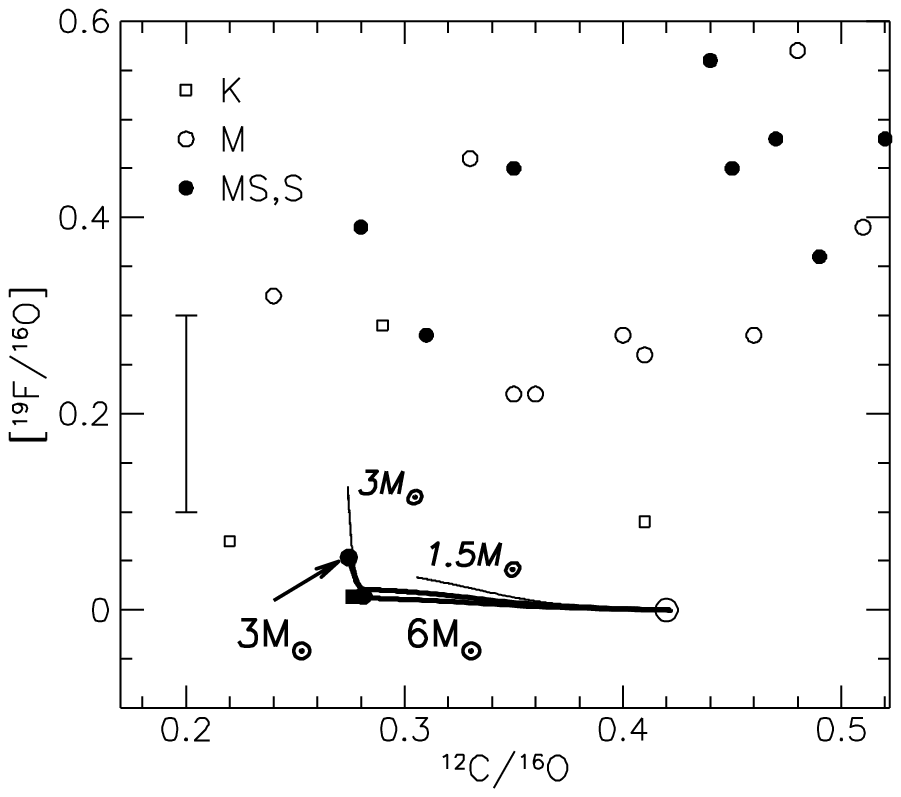}
 \end{picture}
 \vskip -1.8cm
 }
\caption[]{
\label{Fig:FOCOno3dup}
The ([\chem{F}{19}/\chem{O}{16}], \chem{C}{12}/\chem{O}{16}) diagram, comparing
the abundances observed in several classes of red giants from Paper~I with our 
predictions for the first dredge-up (small filled circle along the solid
lines) and the second dredge-up (small filled square along the solid lines).
The thick lines correspond to the M3Z02 and M6Z02 stars, as labeled. The
thin lines correspond to the first dredge-up predictions in 1.5 and
\mass{3} stars (italic labels) when the lower limit is taken for the
\reac{F}{19}{p}{\alpha}{O}{16} rate (Kious 1990), leading to the maximum \chem{F}{19}
production.
Note that, unlike in Paper~I, [\chem{F}{19}/\chem{O}{16}] =
log(\chem{F}{19}/\chem{O}{16})$_*$-log(\chem{F}{19}/\chem{O}{16})$_\odot$,
where subscripts $*$ and $\odot$ refer to stellar and solar abundances,
respectively. The error bar corresponds to the estimated uncertainty on the
observational data
}
\end{figure}

Before considering the fluorine production during the thermally-pulsing
AGB phase, we address the question of surface \chem{F}{19} alterations resulting from
the first and second dredge-ups. 
They occur
shortly after the end of the central hydrogen- and helium-burning phases,
respectively, when the envelope engulfs
the deep layers processed by hydrogen burning (the operation of the second dredge-up being
restricted to stars with $M \simlr$ \mass{4}).

The possibility that the surface fluorine abundance may be altered by the 
first or second dredge-ups is suggested by the observation  
of otherwise normal K and M giants exhibiting
\chem{F}{19}/\chem{O}{16} ratios significantly larger than solar
(Fig.~\ref{Fig:FOCOno3dup}). Since these
stars do not bear the typical signatures of AGB nucleosynthesis (e.g.
overabundances of carbon and elements heavier than iron),
fluorine production in the thermal pulses on the AGB cannot be invoked
to account for their larger-than-solar \chem{F}{19} abundances.

As mentioned in Sect.~\ref{Sect:M6Z02}, fluorine is expected to
be produced as well in hydrogen-burning layers when $T_6<20$ 
(Arnould \& Mowlavi 1993, Arnould et al. 1995), and
destroyed at higher temperatures. This gives rise to a peak in the
\chem{F}{19} profile, as shown in Fig.~\ref{Fig:firstdredgeupprofiles}.
The first dredge-up leads to a moderate increase (by
about 0.05 dex) of the surface \chem{F}{19} abundance in the M3Z02 star,
along with a decrease
of the \chem{C}{12}/\chem{O}{16} ratio (Fig.~\ref{Fig:FOCOno3dup}). No second
dredge-up is experienced by this star.  In the M6Z02
star instead, the temperatures in the hydrogen-burning layers are higher,
resulting in a much thinner \chem{F}{19} peak, so that the surface
\chem{F}{19}/\chem{O}{16} ratio is neither altered after the first,
nor after the second dredge-up.

  Clearly, the first and second dredge-ups in intermediate-mass stars cannot
account for even the lowest F overabundances observed in M giants.
This discrepancy may indicate that:

\noindent (i) the stellar fluorine abundances are systematically overestimated; or

\noindent (ii) the solar-system fluorine abundance is in error; or

\begin{figure}[tbh]
 \hskip -1.3cm
 \parbox[b]{8.0cm}{
 \vskip 0.0cm
 \begin{picture} (8.0,12.1)
   \epsfysize=19.2cm
   \epsfxsize=19.2cm
   \epsfbox{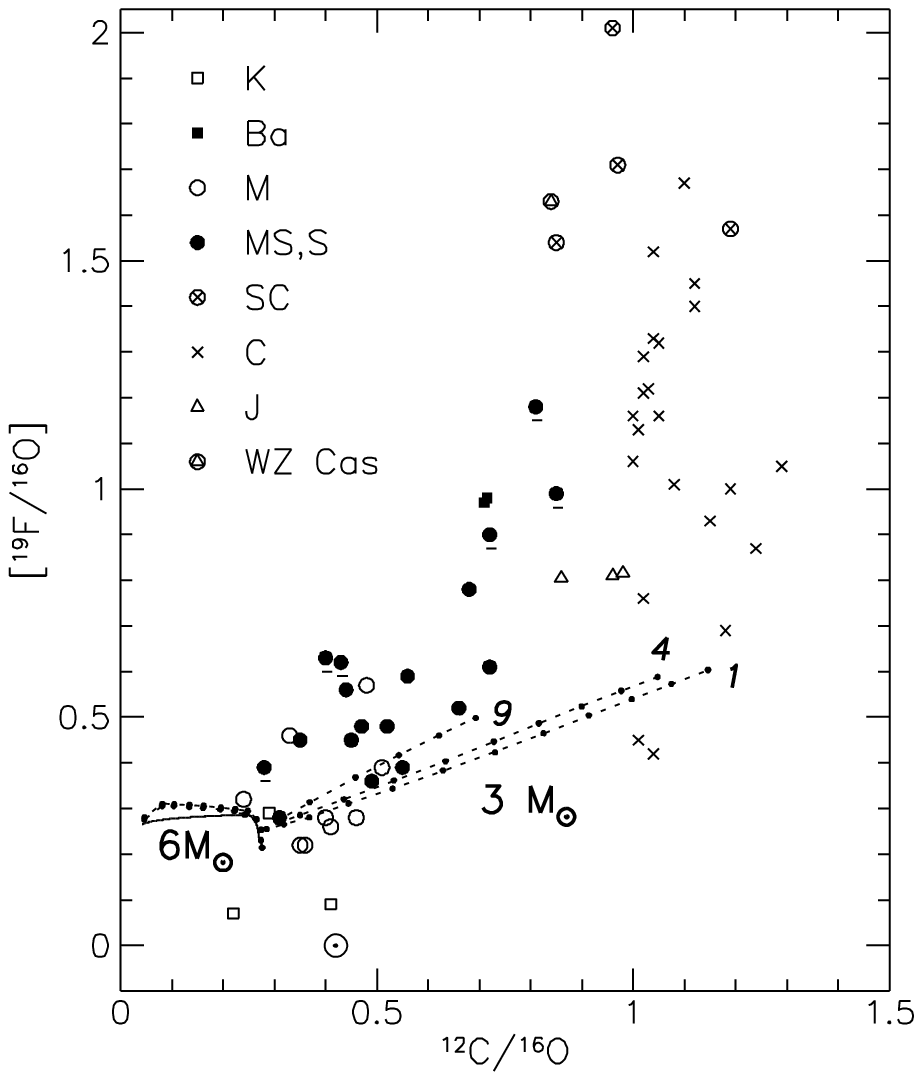}
 \end{picture}
 \vskip -1.9cm
 }
\caption[]{
\label{Fig:FOCO}
Same as Fig.~\protect\ref{Fig:FOCOno3dup}, but for parametrized third dredge-up
predictions. Dotted lines correspond to predictions
with $\lambda=0.3$ and $n_{\mathrm{start}}=1$, 4 or 9, as defined
in the text.  Predictions are normalized in such a way
that the surface F abundances prior to any third dredge-up correspond to the
average  F abundance observed in K and M giants. The solid line corresponds to
the predictions for the M6Z02 star in the absence of third dredge-ups,
reflecting the operation of HBB. Note that time is increasing from left to right
(i.e. increasing C/O) in the M3Z02 tracks, but from right to left (i.e. decreasing
C/O) in the M6Z02 ones.  Underlined symbols denote stars with large N
overabundances (see Paper~I)
}
\end{figure}

\noindent (iii) the actual $\pa{19}{F}{16}{O}$ reaction rate
is slower than the intermediate rate provided
by Kious (1990) and adopted in our calculations.
In fact, that reaction rate is still largely
uncertain in the temperature range of interest  for hydrogen burning
(see Arnould et al. 1995).
To investigate the impact of these uncertainties on the first dredge-up 
predictions, a new evolutionary sequence has been computed up to the
first dredge-up for the \mass{3} star, adopting
the lower limit for the $\pa{19}{F}{16}{O}$ rate.
The surface fluorine overabundance then reaches [F/O] = 0.13~dex
(see Fig.~\ref{Fig:FOCOno3dup}). Although still on the low
side of the observations, these predictions at least fall
within the estimated $\pm 0.1$~dex uncertainty on the observational data.
The case of a solar-metallicity low-mass (\mass{1.5}) star has also been
investigated with the low rate. The resulting surface F enhancement, also
shown in Fig.~\ref{Fig:FOCOno3dup}, is much smaller than in the \mass{3} star.

  Clearly, the elucidation of the differences between the solar-system fluorine abundance
and that of normal K and M giants remains a challenge for future studies. In
Sect.~\ref{Sect:thirddredge-up},
devoted to the analysis of the surface fluorine abundance resulting
from the third dredge-up, we will therefore normalize our predictions in such a way that
the pre-AGB \chem{F}{19} abundance corresponds to the average value observed in
solar-metallicity K and M giants (Fig.~\ref{Fig:FOCO}), 
whatever its origin might be.

\subsection{Third dredge-up}
\label{Sect:thirddredge-up}

  After a pulse, the convective envelope extends inward and, in some cases, can
penetrate the layers formerly involved in the thermal pulse. When this happens,
the He-burning ashes, like \chem{C}{12} and \chem{F}{19}, are mixed in the
convective envelope and brought to the surface. This process, called {\it third
dredge-up}, accounts for the gradual transformation of a M star into a S and
then C star along the AGB.

  Though the very existence of S and C stars on the AGB supports the occurrence
of such dredge-ups, our stellar models like most others fail in obtaining this
process in a self-consistent way. No
fully self-consistent comparison of the observed \chem{F}{19} abundances with
our predictions is thus possible.

  Nevertheless, a comparison is attempted by parametrizing the dredge-ups. Two
parameters need to be specified for that purpose: (i) the depth $\lambda=\Delta
M_d/(M_a+M_b)$, where $\Delta M_d$ is the mass of matter from the pulse mixed
into the convective envelope, and $(M_a+M_b)$ is the mass involved in the
pulse (see Fig.~\ref{Fig:pulseM3Z02}); (ii) the pulse number $n_{\mathrm{start}}$ after
which the third dredge-up first occurs.
It is then assumed that dredge-ups with the same depth $\lambda$ occur after
each subsequent pulses. Though these free parameters introduce some
uncertainties in our predictions, it was shown in Paper I (Eq. 1, Sect. 3.2)
that the use of the (C/O, F/O) diagram strongly limits their impact. Actually,
predictions almost independent of $\lambda$ can be made by using this diagram.

  The parametric dredge-up calculations start with the envelope abundances at
pulse $n_{\mathrm{start}}$ as given by the detailed models. These envelope abundances
are then modified after each subsequent pulse by mixing into the envelope the
matter from the underlying layers, until a fraction $\lambda$ of the ashes of
the pulse has been dredged up. The abundance profiles are taken from the
detailed model calculations. When HBB is taking place, as in the M6Z02 star
(Sect.~\ref{Sect:M6Z02}), its effect on the envelope abundances is incorporated
in the parametric calculations by correcting the abundances by an amount equal
to the relative variation obtained during the interpulse in the detailed
calculations.

  Figure~\ref{Fig:FOCO} displays how the surface (C/O, F/O) ratios vary in
the M3Z02 and M6Z02 stars as a result of the dredge-ups. As discussed in
Sect.~\ref{Sect:firstdredge-up}, our predictions are normalized in such a way
that the surface F abundance prior to any dredge-up corresponds to the average F
abundance observed in K and M giants. Predictions are given for $n_{\mathrm{start}}=1$, 4 and
9, with a dredge-up depth $\lambda=0.3$.
In the case of the M3Z02 star, the surface fluorine overabundances at a given
\chem{C}{12}/\chem{O}{16} ratio are the largest for $n_{\mathrm{start}}=9$
as expected, since the intershell fluorine abundance saturates
from pulse 9 onwards in that star (Fig.~\ref{Fig:N15F19M3Z02}). However, even
in this case, our predictions can only account for the lowest observed fluorine overabundances.
Predictions with $\lambda=0.1$ lead to slopes in the (C/O, F/O) diagram similar
to the ones obtained with $\lambda=0.3$, and are not reported in
Fig.~\ref{Fig:FOCO} for clarity. The above conclusion is thus independent of the
dredge-up parameter $\lambda$.

 In the M6Z02 star, the dredge-ups do not lead to any significant F overabundances.
This is due to the combined effects of the large dilution of the intershell
\chem{F}{19} in the massive envelope, and to HBB. Even though a slight increase
in the surface F abundance is predicted during the early pulses, due mainly to
fluorine production in the envelope itself (see Sect.~\ref{Sect:M6Z02}), the
operation of HBB at later pulses prevents any further increase. Moreover, by
converting the envelope
\chem{C}{12} into \chem{N}{14}, HBB leads to a decrease of the
\chem{C}{12}/\chem{O}{16} ratio in these massive AGB stars. Therefore, it is
rather unlikely that the giants exhibiting large F/O ratios have experienced
HBB, unless their F/O and C/O ratios prior to the onset of HBB were even larger
than their present values. But according to the present models, massive AGB stars
are not very efficient in
increasing their surface F abundance even in the absence of HBB.

  In this respect, WZ~Cas is a puzzling case. This star is known to be super Li-rich
(Denn, Luck \& Lambert 1991, Abia et al. 1993), to have a low
\chem{C}{12}/\chem{C}{13} ratio, and to be enriched in
s-process elements (Dominy 1985). Lithium is produced in the envelope of
massive AGB stars (Sackmann \& Boothroyd 1992) having temperatures $T_6 \simgr
50$ at their base (Mowlavi 1995b). As no large surface F overabundances
are expected in these typical HBB conditions,
the combination of large Li and F excesses at the surface
of WZ~Cas remains unexplained so far. More observations of F abundances in
super Li-rich stars are needed to know whether such a combination is common
among these stars.

  It is noteworthy that J stars (represented by open triangles in
Fig.~\ref{Fig:FOCO}) exhibit rather large F overabundances. This class of
giants is characterized by low \chem{C}{12}/\chem{C}{13} ratios
and no s-process overabundances (Lambert et al. 1986, Dominy 1985). Since their
N abundance is not especially high, HBB alone seems inadequate to account for
their peculiar abundances. The combination of low \chem{C}{12}/\chem{C}{13} and
high \chem{F}{19}/\chem{O}{16} ratios is also puzzling in that respect, as the
extensive hydrogen burning implied by the low \chem{C}{12}/\chem{C}{13} ratio
is also expected to destroy any pre-existing \chem{F}{19}, unless hydrogen
burning occurs at $T_6<20$ (Sect.~\ref{Sect:M6Z02}).

  In conclusion, the surface F abundances predicted for the intermediate-mass
stars considered in this paper are in rather poor agreement with the
observations displayed in Fig.~\ref{Fig:FOCO}. The predicted F abundances in
the M3Z02 star are just able to account for the lowest observed overabundances
(at a given C/O ratio). That conclusion is almost independent of the parameters
$\lambda$ and $n_{\mathrm{start}}$. Actually, a
better match to the observations would require to increase the
\chem{F}{19}/\chem{C}{12} ratio in the pulses by a factor of at least 10 (see
Eq.~1 and Fig.~9 of Paper~I), i.e. $X$(\chem{F}{19}$)\simgr 2\,10^{-4}$ in
the pulse. Equation~(\ref{Eq:F19asymptotic}) indicates that the maximum 
\chem{F}{19} production level in the pulse 
is fixed by the available \chem{C}{13} supply. For secondary \chem{C}{13}
supplied by the CNO cycles operating in the hydrogen-burning shell,  
$X_{\mathrm{a}}(\noy{13}{C}) \sim 10^{-4}$, and the required 
\chem{F}{19} production level is thus not achieved. That conclusion holds true
for low-mass stars.   
It seems thus unavoidable that the most F-rich stars of Fig.~\ref{Fig:FOCO}
require a {\it primary} \chem{C}{13} supply resulting possibly from a
non-standard mixing of protons with the \chem{C}{12} pocket left over by the pulse
(e.g. Hollowell \& Iben 1988). Since a similar need for a primary \chem{C}{13}
supply is in fact coming from the heavy-element overabundances observed in the
same stars (Smith \& Lambert 1990), one may expect the F and s-process
overabundances to be correlated. The fact that the most F-rich stars are also
the most heavy element-rich (Fig.~12 of Paper I) and the most N-rich (at a
given C/O ratio; underlined symbols in Fig.~\ref{Fig:FOCO}) provides evidence
for primary \chem{C}{13} and \chem{N}{14} supplies in these stars.

\begin{figure}[tbh]
 \hskip -1.0cm
 \parbox[b]{8.0cm}{
 \vskip 1.0cm
 \begin{picture} (8.0,8.3)
   \epsfysize=21.5cm
   \epsfxsize=21.5cm
   \epsfbox{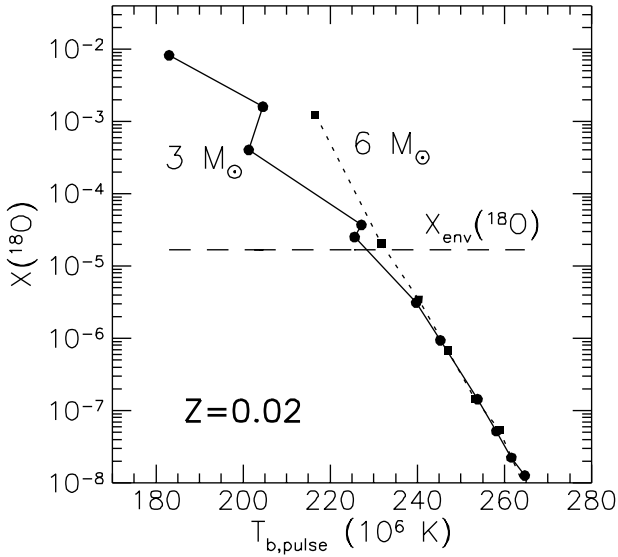}
 \end{picture}
 \vskip -1.7cm
 }
\caption[]{
\label{Fig:O18}
Intershell \chem{O}{18} mass fractions after each pulse as a function of the
maximum pulse temperature for the M3Z02 and M6Z02 stars. The dashed line
corresponds to the surface
\chem{O}{18} abundance of the M3Z02 star prior to any third dredge-up
}
\end{figure}

  Another important constraint on the models comes from the surface
\chem{O}{18} abundance. In Paper~II, large \chem{O}{18} overabundances,
not supported by the observations (Smith \& Lambert 1990), were
predicted to result from the third dredge-up. The new
\reac{O}{18}{\alpha}{\gamma}{Ne}{22} reaction rate (Giesen et al. 1994), which
is faster than the CF88 rate used in Paper~II by a factor of 100 at $T_6=160$,
removes the discrepancy. Figure~\ref{Fig:O18} displays the after-pulse
\chem{O}{18} intershell mass fraction as a function of the maximum pulse
temperature, as well as the envelope abundance prior to any third dredge-up in
the M3Z02 star. It is seen that the after-pulse \chem{O}{18} abundance drops below the
envelope abundance after the 4th pulse in the M3Z02 star. Therefore,
in order to
prevent surface \chem{O}{18} overabundances, third dredge-ups should occur
after the 4th pulse in that star. The \chem{O}{18} overabundance may however be
more difficult to avoid in low-mass stars, where pulses are cooler.

\subsection{Surface fluorine predictions in low-metallicity in\-ter\-me\-dia\-te-mass
            stars}
\label{Sect:lowmetallicity}

\begin{figure}[t]
 \hskip -1.0cm
 \parbox[b]{8.0cm}{
 \vskip 1.1cm
 \begin{picture} (8.0,4.0)
   \epsfysize=18.6cm
   \epsfxsize=18.6cm
   \epsfbox{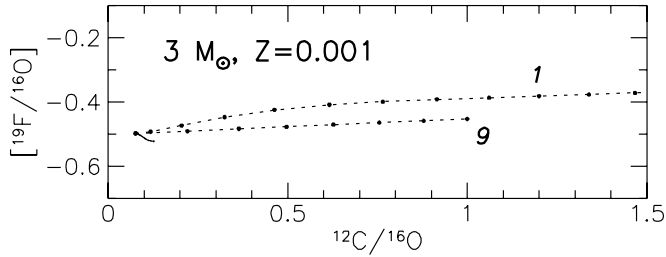}
 \end{picture}
 \vskip -1.9cm
 }
\caption[]{
\label{Fig:FOCOZ001}
Same as Fig.~\protect\ref{Fig:FOCO}, but for M3Z001 (see text). The scales are
the same as in Fig.~\protect\ref{Fig:FOCO}
}
\end{figure}

  Surface fluorine abundances resulting from parametrized third dredge-ups
are predicted for the M3Z001 star in a similar way as for the solar-metallicity
stars (Sect.~\ref{Sect:thirddredge-up}). It is important to note
that these predictions for the low-metallicity star rely on several important
assumptions regarding the {\it initial} envelope abundances. In our models, all
initial abundances are taken equal to the solar-system values, scaled to the
metallicity by the factor Z/Z$_\odot$. Such a scaling rule is justified for C and
supposedly F, since $\mathrm{[C/Fe]=0}$ in unevolved disk stars 
(Edvardsson et al. 1993) and $\mathrm{[F/Fe]\sim 0}$ is suggested  by the observations of
the low-metallicity K giant $\alpha$~Boo (Arcturus, [Fe/H]=\mbox{-0.6}; see Paper~I).
It however does not hold for O, since it is well
known that $\mathrm{[O/Fe]=-0.4\,[Fe/H]}$ in the range $\mathrm{-1\le [Fe/H]\le 0}$ (e.g.
Edvardsson et al. 1993). The initial \chem{O}{16} abundance in the envelope of the
M3Z001 star has been corrected correspondingly. With these assumptions, the
starting point of the dredge-up tracks in the
(\chem{C}{12}/\chem{O}{16},[\chem{F}{19}/\chem{O}{16}]) diagram
corresponds to (0.12,-0.52) (see Fig.~\ref{Fig:FOCOZ001}).

  As can be seen on Fig.~\ref{Fig:FOCOZ001}, third dredge-up episodes in the
M3Z001 star do not alter the surface fluorine abundance as strongly as in the
solar-metallicity stars (Fig.~\ref{Fig:FOCO}). The increase in
[\chem{F}{19}/\chem{O}{16}] with respect to its initial value is only 0.1~dex
after a dozen of pulses, to be compared with as much as 0.4~dex in the 
solar-metallicity stars. 
The less steep slopes predicted in the (C/O, F/O) diagram
for the M3Z001 star result from the combination of two effects: (i) the
lower fluorine production efficiency in the hot pulses of the low-metallicity
star as compared to the ones in the solar-metallicity cases (compare
Figs.~\ref{Fig:N15F19M3Z02} and \ref{Fig:N15F19M3Z001}). The
largest surface F overabundances in the M3Z001 star are therefore obtained 
when the
dredge-ups occur after the early, cool pulses (compare tracks labeled {\it 1}
and {\it 9} in Fig.~\ref{Fig:FOCOZ001}); (ii) the primary nature of \chem{C}{12}
produced in the pulse, as compared to the secondary nature of \chem{F}{19} in
our models. This leads to \chem{F}{19}/\chem{C}{12} ratios in the
dredge-up material lower in low-metallicity stars than in solar-metallicity stars.

  In conclusion, rather small F overabundances are expected in low-metallicity
AGB stars, in the case that 
the fluorine production results from secondary \chem{C}{13}
only.  Observations of F in low-metallicity giants would
definitely be of great interest.

\section{Conclusions}
\label{Sect:Conclusions}

  Detailed calculations of fluorine production have been performed in
intermediate-mass stars with 
(M/M$_\odot$,~Z) = (6,~0.02), (3,~0.02) and (3,~0.001).
It is confirmed that \chem{F}{19} {\it is} produced in pulses through
\reac{N}{14}{\alpha}{\gamma}{F}{18}$(\beta^+)$\reac{O}{18}{p}{\alpha}{N}{15}\reac{}{}{\alpha}{\gamma}{F}{19},
the necessary protons coming from (n,p) reactions with neutrons supplied by
\reac{C}{13}{\alpha}{n}{O}{16}.
Three different regimes of fluorine production can be identified:

\smallskip
\noindent 1) In pulses with maximum base temperatures $T_6\simlr 260$,
\chem{N}{15} is produced {\it within the pulse}.
For $T_6\simlr 220$, \reac{N}{15}{\alpha}{\gamma}{F}{19} is however too slow to
produce significant amounts of F. 
The maximum fluorine production is reached for
$220 \simlr T_6 \simlr 260$,
at a level limited by the \chem{C}{13} supply and the efficiency $g$
of \chem{N}{15} production in the reaction chain starting at \chem{C}{13}.
Typical values for the efficiency are
$g\sim 0.2$, and depend mainly upon the fraction of available 
neutrons involved in (n,p)
reactions. \chem{N}{14} and \chem{Al}{26} are the main neutron-to-proton
converters. The efficiency is found to depend also on the overlap factor
between successive pulses.

\smallskip
\noindent 2) In pulses with $T_6\simgr 260$, the efficiency of \chem{N}{15}
production within the pulse drops dramatically. However, the
temperature of the intershell region now gets high enough for \chem{N}{15} to
be produced {\it in radiative layers} during the interpulse period. When
injected in the next pulse, that \chem{N}{15} is entirely converted into
\chem{F}{19} by $\alpha$-captures. The efficiency of \chem{N}{15} production
is lower in radiative layers than in a pulse
($g_{rad}\sim 0.1$). The intershell \chem{C}{13} nuclei not burned during the
interpulse period are ingested by the pulse close to the time of maximum
pulse temperature. At these high temperatures,
\reac{N}{14}{\alpha}{\gamma}{F}{18} is fast enough to substantially reduce the
number of \chem{N}{14} seeds available to \reac{N}{14}{n}{p}{C}{14}. The
overall efficiency of \chem{F}{19} production is thus lower in this regime than
in regime 1.

\smallskip
\noindent 3) In pulses with $T_6\simgr 300$, \reac{F}{19}{\alpha}{p}{Ne}{22}
destroys \chem{F}{19}.
\smallskip

\medskip
The evolution of the after-pulse fluorine abundance $X_{\mathrm{out}}(\mchem{F}{19})$
in the intershell zone
is dictated by the succession of these three regimes. It first increases during
the first regime until a maximum value is reached, given by $X_{\mathrm{out}}
(\mchem{F}{19})=\frac{19}{13} g X(\mchem{C}{13})$ (where $X(\mchem{C}{13})$
corresponds to the mass fraction of the available \chem{C}{13} supply). During
regime~2, $X_{\mathrm{out}}(\mchem{F}{19})$ drops until it reaches a plateau, with
a level fixed by the efficiency $g_{rad}$. Finally, $X_{\mathrm{out}}
(\mchem{F}{19})$ decreases steadily when regime 3 operates in the late
pulses. {\it The optimum F production in AGB stars is achieved for thermal
pulses having maximum base temperatures in the range $220 \simlr T_6\simlr 260$}.

  All three regimes have been encountered in the M3Z001 star where the pulse
temperatures reach $T_6=260$ at the 4th pulse and 300 at the 13th. The plateau
corresponding to the second regime is reached after 9 pulses in that star. The
M6Z02 star enters regime 2 at the 7th pulse, while the M3Z02 star is still
in regime 1 after 13 pulses.

\medskip
  Comparing our predictions of surface fluorine abundances with the
observations leads to the following conclusions:

\smallskip
\noindent i) Neither the first nor the second dredge-ups
can account for the observed discrepancy between the solar-system
fluorine abundance and that of normal K and M giants.

\smallskip
\noindent ii) Our predictions of surface fluorine abundances resulting from
parametrized third dredge-ups in intermediate-mass stars
can only account for the lowest F overabundances
observed in AGB stars. 
That conclusion is expected to hold true for low-mass stars when fluorine
production results from secondary \chem{C}{13} only.
No concomitant \chem{O}{18} overabundances are
predicted in our intermediate-mass stars, in agreement with the observations.

\smallskip
\noindent iii) The largest F overabundances observed in C stars call for an
additional source of \chem{C}{13}, of {\it primary} origin. A larger \chem{C}{13} supply could
account for the observed overabundances of both F {\it and} s-process elements.
Detailed computations are however necessary to confirm that prediction.

\smallskip
\noindent iv) According to our models, massive AGB stars are not expected to build
up large surface F abundances. Moreover, when operating at $T_6 \ge 70$,
HBB leads to a gradual destruction
of the envelope fluorine, thereby limiting the surface F overabundance that 
would result from the third dredge-up. The large fluorine
overabundance reported for the super Li-rich star WZ~Cas (where HBB
is supposed to be operating) remains therefore unexplained so far.

\smallskip
\noindent v) Hot thermal pulses in intermediate-mass, {\it low-metallicity} AGB
stars are less efficient in producing F. The lower \chem{F}{19}/\chem{C}{12}
ratio from these pulses does not lead to large F overabundances at the surface
of these AGB stars. Any large overabundances that would be
reported by (future) observations of low-metallicity stars, would require
a primary source of \chem{C}{13}.

\appendix
\section*{Appendix: references to nuclear reaction rates}
\label{appendix}

This appendix provides the references for the reaction rates used in the 
present study when different from CF88 or from Beer et al. (1992,
for neutron-capture reactions). Tables~A1, A2 and A3 refer to proton-, 
neutron- and $\alpha$-capture reactions, respectively. The Hauser-Feshbach
method, implemented in the SMOKER code (Thielemann, Arnould \& Truran 1987),
has been used to derive the reaction rates for proton captures 
on \chem{Si}{31}, \chem{Si}{32}, \chem{P}{31},
\chem{P}{32}, \chem{P}{33}, \chem{S}{32}, \chem{S}{33} and \chem{S}{34},
not listed in Table~A1.

\renewcommand{\thetable}{A1}
\begin{table}
\caption[]{\label{Tab:protoncapture}
  References for proton-capture reaction rates, if different from CF88
}
\begin{flushleft}
\begin{tabular}{ll}
\hline\noalign{\smallskip}
\reacbp{C}{11}{p}{\gamma}{N}{12}{C}{12}   &Descouvemont \& Baraffe 1990 \\
\reacbp{N}{13}{p}{\gamma}{O}{14}{N}{14}   &Decrock et al. 1991 \\
\reac{C}{14}{p}{\gamma}{N}{15}            &Wiescher et al. 1990 \\
\reac{O}{17}{p}{\alpha}{N}{14}            &Berheide et al. 1993$^a$ \\
\reac{O}{17}{p}{\gamma}{F}{18}            &Berheide et al. 1993$^a$ \\
\reac{F}{19}{p}{\alpha}{O}{16}            &Kious 1990 \\
\reac{Ne}{21}{p}{\gamma}{Na}{22}          &G\"orres et al. 1982, G\"orres\\
                                          &\hskip 1mm et al. 1983 \\
\reac{Ne}{22}{p}{\gamma}{Na}{23}          &G\"orres et al. 1983 \\
\reacbp{Na}{22}{p}{\gamma}{Mg}{23}{Na}{23}&Seuthe et al. 1990$^b$ \\
\reac{Na}{23}{p}{\gamma}{Mg}{24}          &CF88 + G\"orres et al. 
                                           1989$^c$ \\
\reac{Mg}{25}{p}{\gamma}{Al^g}{26}        &Iliadis et al. 1990$^d$ \\
\reac{Mg}{25}{p}{\gamma}{Al^i}{26}        &Iliadis et al. 1990$^d$ \\
\reac{Mg}{26}{p}{\gamma}{Al}{27}          &Iliadis et al. 1990$^d$ \\
\reacbp{Al^g}{26}{p}{\gamma}{Si}{27}{Al}{27}&Champagne et al. 1993$^a$ \\
\reac{Al}{27}{p}{\gamma}{Si}{28}          &Timmermann et al. 1988$^a$ \\
\reac{Al}{27}{p}{\alpha}{Mg}{24}          &Timmermann et al. 1988 \\
                                          &\hskip 1mm + Champagne et al. 1988$^{e,a}$\\
\reacbp{Si}{28}{p}{\gamma}{P}{29}{Si}{29} &Graff et al. 1990 \\
\reacbp{S}{32}{p}{\gamma}{Cl}{33}{S}{33}  &Iliadis et al. 1992 \\
\noalign{\smallskip}
\hline
\end{tabular}
 \\
 $^a$ Geometric average between lower and upper limits\\
 $^b$ Resonance at 70 keV from Berheide et al. (1993), plugged in 
expression from Landr\'e et al. (1989) with f${_1}$=f${_2}$=1\\
 $^c$ Resonant contributions from G\"orres et al. (1989)\\
 $^d$ Geometric average between lower and upper limits for $T_9>0.01$\\
 $^e$ Resonant contributions from Champagne et al. (1988)\\
\end{flushleft}
\end{table}

\renewcommand{\thetable}{A2}
\begin{table}
\caption[]{\label{Tab:ncapture}
  References for neutron-capture reaction rates, if different from Beer, 
Voss \& Winters (1992)
}
\begin{flushleft}
\begin{tabular}{l l}
\hline\noalign{\smallskip}
\reac{C}{12}{n}{\gamma}{C}{13}                & Ohsaki et al. 1994 \\
\reac{C}{13}{n}{\gamma}{C}{14}                & Raman et al. 1990 \\
\reac{N}{13}{n}{p}{C}{13}                     & CF88 (reverse) \\
\reac{N}{14}{n}{p}{C}{14}                     & Koehler \& O'Brien 1989\\
\reacbm{N}{15}{n}{\gamma}{N}{16}{O}{16}       & FCZ67\\
\reac{O}{15}{n}{p}{N}{15}                     & CF88 (reverse) \\
\reac{O}{17}{n}{\gamma}{O}{18}                & Wagoner 1969 \\
\reac{O}{17}{n}{\alpha}{C}{14}                & Schatz et al. 1993 \\
\reacbm{O}{18}{n}{\gamma}{O}{19}{F}{19}       & Rauscher et al. 1994 \\
\reac{F}{18}{n}{\alpha}{N}{15}                & Thibaud 1989 \\
\reac{F}{18}{n}{p}{O}{18}                     & Thibaud 1989 \\
\reac{Ne}{21}{n}{\gamma}{Ne}{22}              & Almeida \& K\"appeler 1983 \\
\reac{Al^g}{26}{n}{\gamma}{Al}{27}            & WFHZ78 \\
\reac{Al^g}{26}{n}{p}{Mg}{26}                 & CF88 \\
\reac{Al^g}{26}{n}{\alpha}{Na}{23}            & CF88 \\
\reac{Al^i}{26}{n}{\gamma}{Al}{27}            & WFHZ78 \\
\reac{Al^i}{26}{n}{\alpha}{Na}{23}            & CF88 \\
\reac{Al^i}{26}{n}{p}{Mg}{26}                 & CF88 \\
\reac{Si}{31}{n}{\gamma}{Si}{32}              & SMOKER \\
\reacbm{Si}{32}{n}{\gamma}{Si}{33}{P}{33}     & SMOKER \\
\reac{P}{32}{n}{\gamma}{P}{33}                & SMOKER \\
\reac{P}{32}{n}{p}{Si}{32}                    & SMOKER \\
\reac{P}{33}{n}{\beta^-}{S}{34}               & SMOKER \\
\reac{S}{33}{n}{p}{P}{33}                     & SMOKER \\
\reac{S}{33}{n}{\alpha}{Si}{30}               & Wagemans et al. 1987 \\
\noalign{\smallskip}
\hline
\end{tabular}
\end{flushleft}
\end{table}

\renewcommand{\thetable}{A3}
\begin{table}
\caption[]{\label{Tab:alphacapture}
  References for $\alpha$-capture reaction rates, if different from CF88
  }
\begin{flushleft}
\begin{tabular}{l p{5cm}}
\hline\noalign{\smallskip}
\reac{C}{12}{\alpha}{\gamma}{O}{16}           & Caughlan et al. 1985 \\
\reac{C}{13}{\alpha}{n}{O}{16}                & Drotleff et al. 1993 \\
\reac{C}{14}{\alpha}{\gamma}{O}{18}           & Hashimoto et al. 1986, 
                                                Gai et al. 1987, 
                                                Funck \& Langanke 1989 \\
\reac{N}{15}{\alpha}{\gamma}{F}{19}           & de Oliveira et al. 1995 \\
\reac{O}{18}{\alpha}{\gamma}{Ne}{22}          & Giesen et al. 1994 \\
\reac{Ne}{22}{\alpha}{n}{Mg}{25}              & Drotleff et al. 1993 \\
\reac{Ne}{22}{\alpha}{\gamma}{Mg}{26}         & Drotleff et al. 1992, priv. comm. \\
\noalign{\smallskip}
\hline
\end{tabular}
\end{flushleft}
\end{table}

\vskip 5mm
{\noindent\bf References to tables}

\footnotesize{

\ind
Almeida J., K\"appeler F., 1983, ApJ 265, 417

\ind
Berheide M., Rolfs C., Schroeder U., Trautvetter H.P., 1993, Z.Phys.A 343, 483

\ind
Caughlan G.R., Fowler W.A., Harris M.J., Zimmerman B.A., 1985,
Atom. Data Nucl. Data Tables 32, 197

\ind
Champagne A.E., Cella C.H., Kouzes R.T., Lowry M.M., Magnus P.V., Smith M.S., 
Mao Z.Q., 1988, Nucl. Phys. A487, 433 

\ind
Champagne A.E., Brown B.A., Sherr R., 1993, Nucl.Phys. A556, 123

\ind
Decrock P., Delbar Th., Duhamel P., Galster W., 
Huyse M., Leleux P., Licot I., Li\'enard E., 
Lipnik P., Loiselet M., Michotte C., Ryckewaert G., 
Van Duppen P., Vanhorenbeeck J., Vervier J., 1991, 
Phys. Rev. Lett. 67, 808

\ind
Descouvemont P., Baraffe I., 1990, Nucl.Phys. A514, 66

\ind
de Oliveira F., Coc A., Aguer P., Angulo C.,
Bogaert G., Kiener J., Lefebvre A.,
Tatischeff V., Thibaud J.P., Fortier S.,
Maison J.M., Rosier L., Rotbard G., Vernotte J.,
Arnould M., Jorissen A., Mowlavi N., 1995, submitted to Nucl. Phys.

\ind
Drotleff H.W., Denker A., Knee H., Soine M., Wolf G.,
Hammer J.W., Greife U., Rolfs C., Trautvetter H.P., 1993, ApJ 414, 735

\ind
Fowler W.A., Caughlan G.R., Zimmerman B.A., 1967, ARA\&A 5, 525 (FCZ67)

\ind
Funck C., Langanke K., 1989, ApJ 344, 46 

\ind
Gai M., Keddy R., Bromley D.A., Olness J.W., 
Warburton E.K., 1987, Phys. Rev. C36, 1256

\ind
Giesen U., Browne C.P., G\"orres J., Ross J.G., 
Wiescher M., Azuma R.E., King J.D., Vise J.B., 
Buckby M., 1994, Nucl. Phys. A567, 146

\ind
G\"orres J., Rolfs C., Schmalbrock P., Trautvetter H.P., Keinonen J., 
1982, Nucl.Phys. A385, 57

\ind
G\"orres J., Becker H.W., Buchmann L., Rolfs C., 
Schmalbrock P., Trautvetter H.P., Vlieks A., 
Hammer J.W., Donoghue T.R., 1983, Nucl.Phys. A408, 372

\ind
G\"orres J., Wiescher M., Rolfs C., 1989, ApJ 343, 365

\ind
Graff S., G\"orres J., Wiescher M., Azuma R.E., King J., 
Vise J., Hardie G.H., Wang T.R., 1990, Nucl. Phys. A510, 346

\ind
Hashimoto M., Nomoto K., Arai K., Kaminisi K., 1986, ApJ 307, 687

\ind
Iliadis Ch., Schange Th., Rolfs C., Schroeder U., 
Somorjai E., Trautvetter H.P., Wolke K., Endt P.M., 
Kikstra S.W., Champagne A.E., Arnould M., Paulus G., 
1990, Nucl.Phys. A512, 509

\ind
Iliadis C., Giesen U., G\"orres J., Wiescher M.,
Graff S.M., Azuma R.E., Barnes C.A., 1992, Nucl.Phys. A539, 97

\ind
Kious 1990, Ph.D. thesis, Orsay

\ind
Koehler P.E., O'Brien 1989, Phys. Rev. C39, 1655

\ind
Landr\'e V., Aguer P., Bogaert G., Lefebvre A., Thibaud J.P., Fortier S., 
Maison J.M., Vernotte J., 1989, Phys.Rev. C40, 1972

\ind
Ohsaki T., Nagai Y., Igashira M., Shima T., Takeda K., Seino S., Irie T., 
1994, ApJ 422, 912

\ind
Raman S., Igashira M., Dozono Y., Kitazawa H., 
Mizumoto M., Lynn J.E., 1990, Phys.Rev. C41, 458

\ind
Rauscher Th., Applegate J.H., Cowan J.J., 
Thielemann F.-K., Wiescher M., 1994, ApJ 429, 499

\ind
Schatz H., K\"appeler F., Koehler P.E.,
Wiescher M., Trautvetter H.P., 1993, ApJ 413, 750

\ind
Seuthe S., Rolfs C., Schroeder U., Schulte W.H.,
Somorjai E., Trautvetter H.P., Waanders, F.B., 
Kavanagh R.W., Ravn H., Arnould M., Paulus G., 
1990, Nucl. Phys. A514, 471

\ind
Thibaud J.P., 1989, personal comm.

\ind
Thielemann F.-K., Arnould M., Truran J.W., 1986. In: Vangioni-Flam E., 
Audouze J., Cass\'e M.,
Chi\`eze J.P., Tran Thanh Van J. (eds.) Advances in 
Nuclear Astrophysics. Ed. Fronti\`eres,
Gif-sur-Yvette, p. 525 

\ind
Timmermann R., Becker H.W., Rolfs C., Schroeder U., 
Trautvetter H.P., 1988, Nucl.Phys. A477, 105

\ind
Wagemans C., Weigmann H., Barthelemy R., 1987, Nucl.Phys. A469, 497

\ind
Wagoner R.W., 1969, ApJS 18, 247

\ind
Wiescher M., G\"orres J., Thielemann F.-K., 1990, ApJ 363, 340

\ind 
Woosley S.E., Fowler W.A., Holmes J.A., 
Zimmerman B.A., 1978, Atomic Data Nucl. Data Tables 22, 371 (WFHZ78)

}


\begin{thebibliography}{}

\bibitem[]{}
Abia C., Boffin H.M.J., Isern J., Rebolo R., 1993, A\&A 272, 455

\bibitem[]{}
Arnould  M., Mowlavi N., 1993. In: Weiss W. W., Baglin, A. (eds.)
Inside the Stars. (IAU Coll. 137). ASP Conf. Ser. 40, p. 310

\bibitem[]{}
Arnould  M., Mowlavi N., Champagne A., 1995. In: Stellar Evolution: What Should Be
Done; $32^{\mathrm{nd}}$ Li\`ege Int. Astroph.  Coll. 1995, in press

\bibitem[]{}
Beer H., Voss F., Winters R.R., 1992, ApJS 80, 403

\bibitem[]{}
Caughlan  G.R., Fowler  W.A., 1988, Atomic Data and Nucl. Data Tables  40, 283
(CF88)

\bibitem[]{}
Charbonnel  C., 1994, A\&A 282, 811

\bibitem[]{}
Denn G.R., Luck R.E., Lambert D.L., 1991, ApJ 377, 657

\bibitem[]{}
de Oliveira F., Coc A., Aguer P., Angulo C., Bogaert G., Kiener J., 
Lefebvre A., Tatischeff V., Thibaud J.P., Fortier S.,
Maison J.M., Rosier L., Rotbard G., Vernotte J.,
Arnould M., Jorissen A., Mowlavi N., 1995, Nucl.Phys., in press

\bibitem[]{}
Dominy J.F., 1985, PASP 97, 1104

\bibitem[]{}
Edvardsson B., Andersen J., Gustafsson B., Lambert D.L., Nissen P.E., Tomkin J., 1993,
A\&A 252, 597 

\bibitem[]{}
Forestini M., Paulus G., Arnould M., 1991, A\&A 252, 597 

\bibitem[]{}
Forestini M., Goriely S., Jorissen A., Arnould M., 1992, A\&A 261, 157 (Paper~II)

\bibitem[]{}
Giesen U., Browne C.P., G\"orres J., Ross J.G.,
Wiescher M., Azuma R.E., King J.D., Vise J.B., 
Buckby M., 1994, Nucl. Phys. A567, 146

\bibitem[]{} 
Gilroy K.K., Brown J.A., 1991, ApJ 371, 578

\bibitem[]{} 
Hollowell D., Iben I., 1988, ApJ 333, L25

\bibitem[]{} 
Jorissen A., Smith V.V., Lambert D.L., 1992, A\&A 261, 164 (Paper I)

\bibitem[]{} 
Lambert D.L., Gustafsson B., Eriksson K., Hinkle K.H., 1986, ApJS 62, 373

\bibitem[]{}
Mowlavi N., 1995a, Ph.D. thesis, Universit\'e Libre de Bruxelles  

\bibitem[]{}
Mowlavi N., 1995b. In: Crane Ph. (ed.) Proc. ESO/EIPC Workshop,
The Light Element Abundances. Springer, p. 297

\bibitem[]{} 
Prantzos N., Arnould M., Arcoragi J.P., 1987, ApJ 315, 209

\bibitem[]{} 
Sackmann I.-J., Boothroyd A.I., 1992, ApJ 392, L71

\bibitem[]{} 
Smith V.V., Lambert D.L. 1990, ApJS, 72, 387

\bibitem[]{} 
Wallerstein G., Morell O., 1994, A\&A 281, L37

\bibitem[]{} 
Ziurys L. M., Apponi A. J., Phillips T. G., 1994, ApJ 433, 729

\end{thebibliography}
\end{document}